\documentclass[conference]{IEEEtran}
\IEEEoverridecommandlockouts
\usepackage{cite}
\usepackage{amsmath,amssymb,amsfonts}
\usepackage{algorithmic}
\usepackage{graphicx}
\usepackage{textcomp}
\usepackage{xcolor}

\usepackage{orcidlink}
\usepackage{url}
\usepackage[caption=false,font=normalsize,labelfont={scriptsize},textfont=sf]{subfig}
\usepackage{stfloats}
\usepackage[none]{hyphenat}
\usepackage{paralist}
\usepackage[detect-all]{siunitx}
\def\BibTeX{{\rm B\kern-.05em{\sc i\kern-.025em b}\kern-.08em
    T\kern-.1667em\lower.7ex\hbox{E}\kern-.125emX}}
\begin{document}

\title{InfiniteEn: A Multi-Source Energy Harvesting System with Load Monitoring Module  for Batteryless Internet of Things}

\author{\IEEEauthorblockN{ Priyesh Pappinisseri Puluckul}
\IEEEauthorblockA{\textit{IDLab—Faculty of Applied Engineering, } \\
\textit{University of Antwerp—imec,}\\
Antwerp, Belgium \\
priyesh.pappinisseripuluckul@uantwerpen.be \orcidlink{0000-0003-4145-9443} }
\and
\IEEEauthorblockN{Maarten Weyn}
\IEEEauthorblockA{\textit{IDLab—Faculty of Applied Engineering, } \\
\textit{University of Antwerp—imec,}\\
Antwerp, Belgium \\
maarten.weyn@uantwerpen.be \orcidlink{0000-0003-1152-6617} }
}

\maketitle

\begin{abstract}
This paper presents  InfiniteEn,  a multi-source energy harvesting platform designed for the Internet of Batteryless Things (IoBT). InfiniteEn incorporates an efficient energy combiner to combine energy from different harvesting sources. The energy combiner uses capacitor-to-capacitor energy transfer to combine energy from multiple sources and achieves a nominal efficiency of 88\%. In addition to multiplexing different sources, the energy combiner facilitates the estimation of the harvesting rate and the calibration of the capacity of the energy buffer. The energy storage architecture of InfiniteEn employs an array of storage buffers that can be configured on demand to cope with varying energy harvesting rates and load's energy requirements. To address the challenge of tracking the energy state of batteryless devices with minimum energy overhead, this work introduces the concept of a Load Monitoring Module (LMM).  InfiniteEn is a load-agnostic platform, meaning that it does not require any prior knowledge of the energy profile of the load to track its energy states. The LMM  assists InfiniteEn in tracking the energy state of the load and dynamically modifying the storage buffers to meet the load's energy requirements.  Furthermore, the module can detect and signal any abnormalities in the energy consumption pattern of the load caused by a hardware or software defect. Experiments demonstrate that LMM has a response time of less than  \SI{11}{\milli\second} to energy state changes.
\end{abstract}


\section{Introduction}

\IEEEPARstart{I}{nternet} of Batteryless Things (IoBT) are envisioned as a step forward to a sustainable and environment-friendly Internet of Things (IoT). Batteryless devices are energy-neutral devices which achieve energy autonomy by scavenging energy from environmental sources such as solar, heat, radio waves, etc. They eliminate batteries and store collected energy in alternate energy buffers like supercapacitors  which have considerably higher lifetimes  and are more environmentally friendly compared to batteries~\cite{luxbeacon, camaroptera}. Despite providing an efficient and sustainable solution to overcome the limitations of battery-based IoT, IoBT still faces many challenges that hinder its extensive adoption. Energy harvesting can be inconsistent and irregular~\cite{future_of_sensing}. Moving from a battery, which has a high energy density, to a low energy density system like a supercapacitor requires optimal utilization of the energy to avoid device blackout in case of an energy outage. Moreover, the usage of supercapacitors may lead to frequent power failures of the system when the energy sources are weak. So, to mitigate the impact of energy outages, it is essential that the system has an idea of both the available energy and its own energy consumption. 

\begin{figure}	

\centering
\captionsetup[subfloat]{labelformat=empty}
 \subfloat[]{%
     \includegraphics[width=1.6in]{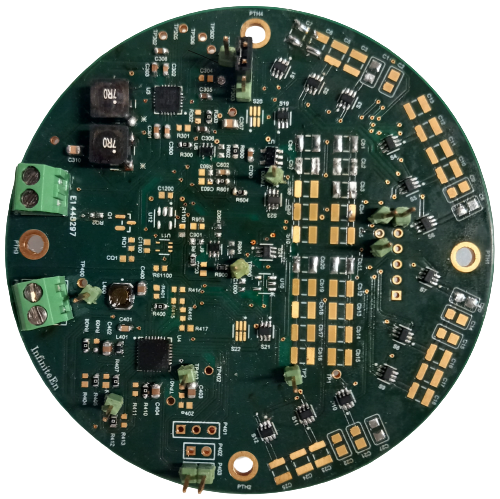}
     \label{fig_lpcb_top}
    }    
    \quad 
 \subfloat[]{%
     \includegraphics[width=1.6in]{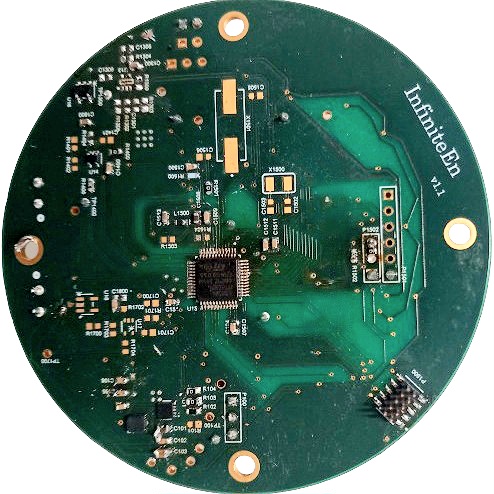}
      \label{fig_lpcb_bot}
   }    
\caption{A photo   of InfiniteEn designed and fabricated  on a   PCB of \SI{60}{\milli\meter} diameter.  \protect\subref{fig_lpcb_top} top side \protect\subref{fig_lpcb_bot} bottom side.}
\label{fig_infiniteen_board}
\vspace{-5mm}
\end{figure}

\begin{figure*}
\includegraphics[width=5.0in]{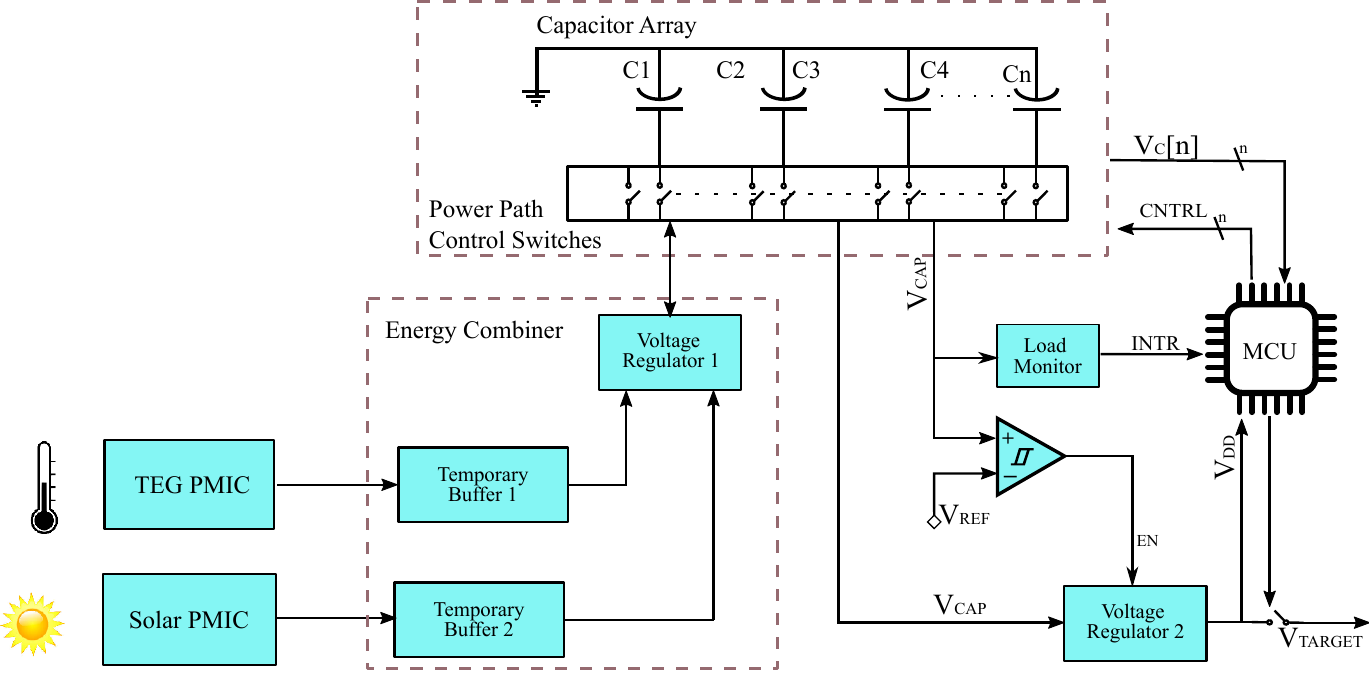}
\centering
\caption{A simplified block diagram representation  of the InfiniteEn architecture.}
\label{fig_high_level_architecture}
\vspace{-2mm}
\end{figure*}

The reliability of a batteryless device depends on the available energy budget and optimum utilization of the energy. Consequently,   batteryless designs started using real-time schedulers that can manage task execution based on the energy budget of the system\cite{optimal_energy_aware_scheduler,energy_aware_scheduler}.  Real-time schedulers can help to optimize task execution based on the energy requirement of the tasks and energy availability from ambient sources. Though real-time schedulers improve the performance of IoBT, they often require knowledge of energy availability and task-wise energy consumption of the load. The majority of these solutions use empirically acquired values of energy consumption in their design~\cite{optimal_energy_aware_scheduler, energy_aware_scheduler }. Real-time schedulers are software enteric approaches and ensure the efficient use of stored energy by optimizing the task executions. Instead, re-configurable storage architectures try to optimize the size of the storage buffers based on the load requirement and energy availability~\cite{capybara, morphy, architectural_charge_management}. Whereas, multisource energy harvesting systems try to increase the energy budget by scavenging energy from more than one sources~\cite{bg_multisource_harvesting, eneragy_combiner_multi_input_boost, energy_combiner_single_inductor_boost}.

 
Reliable batteryless designs demand more advanced hardware platforms in combination with energy-optimized software modules. In this paper, we introduce InfiniteEn, a multisource  energy harvesting system for batteryless IoT devices and present its discussion and evaluation in detail. InfiniteEn addresses the lack of a hardware platform which can exploit multiple sources of energy and can monitor the state of both the load and energy sources in realtime.  Unlike many existing systems where less efficient energy OR-ing circuits are employed to combine energy from multiple source~\cite{energy_combiner_oring1, energy_combiner_oring2}, InfiniteEn uses \textit{capacitor-to-capacitor} energy transfer circuit to harvest energy from multiple sources simultaneously. Further to combining energy, the energy combiner can also measure the harvesting rate and energy transferred to the supercapacitor. Though the energy combiner used with infiniteEn can theoretically work with any number of sources, we consider only solar and thermal energy sources in this paper.

infiniteEn employs a dynamic energy storage architecture similar to Capybara and Morphy~\cite{capybara, morphy}. However,  InfiniteEn does not require any prior details of the energy profile of the load to dynamically adjust the energy buffer. Instead, infiniteEn incorporates a novel circuit called a Load Monitoring Module (LMM) which can monitor the energy state of the target device.  With the help of LMM, InfiniteEn releases programmers from the burden of precisely measuring the energy consumption of each application task, which is often the case with existing solutions \cite{optimal_energy_aware_scheduler, energy_aware_scheduler, morphy, capybara}.  The device could directly use the energy state information available from the LMM to dynamically adjust the storage buffers on the run. In addition to this, by following the energy states, LMM can also detect and flag any defect or bug in the target code or hardware that may lead to unwanted energy leakages.

In general, we make the following contributions in this paper,

\begin{itemize}
    \item We present the design and evaluation of InfiniteEn, a multi-source energy harvesting architecture with energy awareness for batteryless IoT devices. 

    \item We  design  an \textit{energy combiner} to combine energy from different sources  based on the concept of \textit{capacitor-to-capacitor energy transfer}.  The energy combiner can further assist in  estimating the rate of energy harvesting and calibrating the capacitance of the storage buffers.
      
    \item We introduce a novel concept and design of \textit{LMM}  which can track the energy state of the application program with minimum energy overhead. The LMM can assist the system in deciding the capacity of the storage buffer required.

\end{itemize}
\vspace{0.2mm}
The remainder of this paper is organised as follows: A detailed discussion on the hardware architecture of InfiniteEn is presented in Section \ref{section_infiniteEn_architecture}.  We further present the evaluation of InfiniteEn and its different components in Section \ref{section_evaluation}.  Finally, conclusions are presented in Section \ref{section_conclusion}.

\section{InfiniteEn Architecture }
\label{section_infiniteEn_architecture}

A simplified block diagram of  InifinteEn is shown in Fig.~\ref{fig_high_level_architecture}. The system consists of  Power Management Integrated Circuits (PMIC) for solar and thermal energy harvesting, an array of supercapacitors that can be adjusted on the fly, a    capacitor-based energy combiner, the LMM and a Microcontroller that manages InifiniteEn states. 
 
\subsection{Supercapacitor Array}
\begin{figure}
\centering
\includegraphics[width=2.5in]{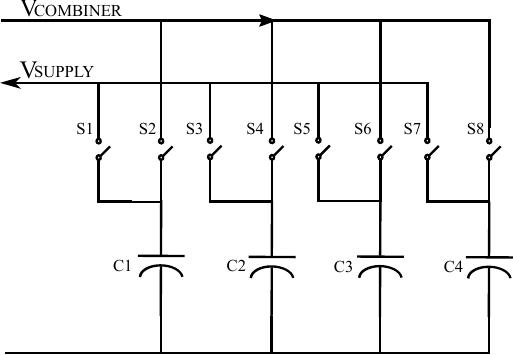}
\caption{Capacitor array with re-configurable charge-discharge path.}
\label{fig_capacitor_array}
\vspace{-5mm}
\end{figure}

The supercapacitor array consists of a group of supercapacitors that can be adjusted on demand to make different parallel combinations of capacitors. As in Capybara~\cite{capybara}, InfiniteEn uses the capacitor array to deal with the energy harvesting rate and to allocate energy for the application code. A schematic of the capacitor array is shown in Fig.~\ref{fig_capacitor_array}. The  capacitors can be arranged in different parallel connections using the low power switches. All the capacitors are independent, and multiple capacitors can be connected to the input of the  voltage regulator ($V_{SUPPLY}$) or to the output of the energy combiner ($V_{COMBINER}$) simultaneously. Currently,  an array of 4 capacitors with capacitance \SI{15}{\milli\farad}, \SI{33}{\milli\farad}, \SI{68}{\milli\farad} and \SI{100}{\milli\farad} is used, giving a total capacitance of \SI{216}{\milli\farad} \cite{bestcap}.

\subsection{Power Management Circuit}
The Power management circuit for the energy harvesters consists of  two PMICs, one for solar energy and the other for thermal energy. We use AEM10491 \cite{aem10491} from e-peas for solar energy harvesting  and LTC3109 \cite{ltc3109} from Analog Semiconductors for thermal energy harvesting. AEM10491 can harvest energy from sources as low as \SI{50}{\milli\volt}, whereas LTC3109 can work with Thermal Energy Generator (TEG) output as low as \SI{30}{\milli\volt}. Since the voltage from the capacitors varies with time, it is essential to add a voltage regulator circuit before feeding to the other components on the board. We choose TPS63900  from Texas Instruments for this purpose~\cite{tps63900}.  TPS63900 can regulate any voltage between \SI{1.8}{\volt} and \SI{5.5}{\volt} while consuming only \SI{75}{\nano\ampere}.



\subsection{Energy Combiner and Energy Estimation}

\begin{figure}[h]
\centering
\includegraphics[width=3.0in]{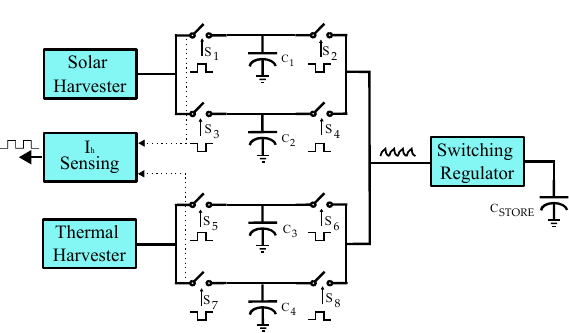}
\caption{A schematic representation of energy combiner circuit with both thermal and solar energy harvesting source.}
\label{fig_energy_combiner}
\vspace{-5mm}
\end{figure}

 InfiniteEn achieves high efficiency in energy addition with the help of a \textit{capacitor-to-capacitor energy transfer} circuit. The energy combiner design is inspired by \textit{Energy Bucket} \cite{energy_bucket}. A simplified representation of the energy combiner is shown in Fig.~\ref{fig_energy_combiner}.   

\begin{figure}
\centering
\includegraphics[width=2.0in]{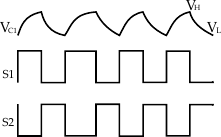}
\caption{A schematic representation of the operation of energy combiner showing the status of switches $S_{1}, S_{2}, S_{3}, S_{4}$ and the voltage $V_{CAP}$ at the input of the switching regulator.}
\label{fig_energy_combiner_timing_diagram}
\vspace{-6mm}
\end{figure}

The circuit employs 4 capacitors ($C_{1}$-$C_{4}$), two per each energy harvesting source, eight switches ($S_{1}$-$S_{8}$), and a switching regulator as shown in Fig.~\ref{fig_energy_combiner}. The capacitors are used as temporary buffers to hold energy for a short time before it gets transferred to the main storage $C_{Store}$ via a switching voltage regulator.  SiP323432(1) is used for controlling the charge-discharge paths, and TPS63900 is again used as the voltage regulator.  Tantalum capacitors are chosen for temporary buffers since their capacitance is independent of the bias voltage.  The switches alternate the connection of the capacitors between the energy source and the switching regulator, creating a charge-discharge cycle as shown in Fig.~\ref{fig_energy_combiner_timing_diagram}. For instance, when a switch $S_{1}$ is ON, the corresponding  capacitor, $C_{1}$ gets connected to the source. Once the capacitor gets connected, the source  charges it to a voltage level $V_{H}$. Upon reaching $V_{H}$, $S_{1}$ is turned OFF and $S_{2}$ is activated. At the same time, $S_{3}$ is turned ON so that $C_{2}$ gets charged while $C_{1}$ is unavailable.  On activating $S_{2}$,   the switching regulator transfers the energy stored in $C_{1}$ to the main capacitor $C_{STORE}$ until the voltage across it reaches the lower threshold $V_{L}$. This process repeats with each cycle transferring $E_{cap}=\frac{1}{2}C(V_{H}^2 - V_{L}^2)$ of energy from temporary buffers to the main buffer. o, the total energy transferred to the main storage  can be calculated by counting the number of threshold crossings  as,


\begin{equation}
\label{eqn_stored_energy}
E_{STORE} =\eta \sum_{i=0}^{n} { E_{cap}}
\end{equation}

 Where $\eta$ is the efficiency of the voltage regulator. The efficiency of the regulator can be either extracted from the datasheet or can be determined empirically.  Alternatively, $\eta E_{cap}$ which is the energy transferred per cycle, can be measured with a power analyzer like Joulescope \cite{joulescope}. 

The energy combiner circuit can  further measure the harvesting  rate by counting the charging-discharging cycles of the temporary capacitors. This is possible since the charging time of the temporary buffer is directly proportional to the harvesting rate. Thus the energy harvesting rate is given as, 

\begin{equation}
    I_{harvest} = C.N\frac{(v_{H}-v_{L})}{T}
    \label{eqn_average_harvester_current}
\end{equation}

Where $N$ is the number of charging-discharging cycles in a period of $T$ seconds.

\subsection{Load Monitoring Module}
\begin{figure}[h]
\centering
\includegraphics[width=3.5in]{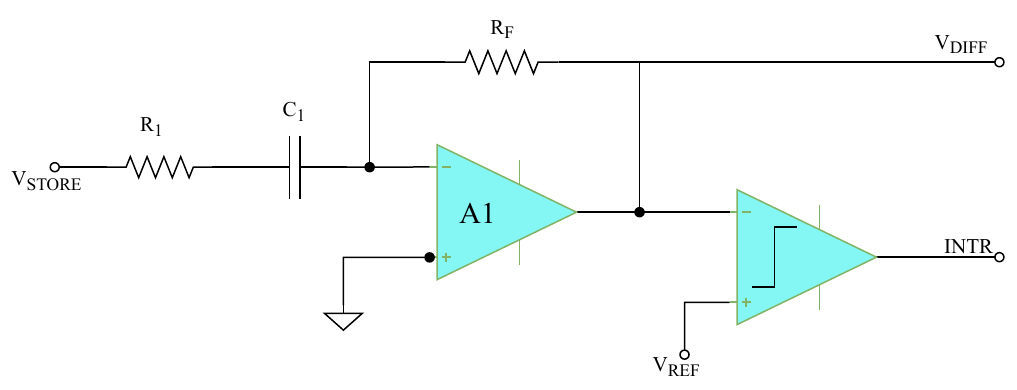}
\caption{A schematic representation of the load monitoring module. The module uses an analogue differentiator and a comparator circuit to detect changes in $V_{STORE}$.}
\label{fig_load_monitor_circuit}
\end{figure}
Monitoring and tracking the energy consumption of the target device is a crucial requirement in a batteryless system design. Any abnormal activities of the load can lead to the failure of the system. An abnormal activity can be defined as any activity where the energy demand is disproportional to the expected behaviour or to the available energy. LMM, a novel concept integrated into infiniteEn  can efficiently monitor the energy state of the supercapacitor. A simple analogue differentiator with a comparator circuit as shown in Fig.~\ref{fig_load_monitor_circuit} is used for this purpose. The differentiator produces a voltage corresponding to the $\frac{dv}{dt}$ of the supercapacitor, which is further converted into an interrupt signal by the comparator.  For a capacitor, the current drawn, $I_{cap} = C\frac{dv}{dt}$, which means that the output of the differentiator circuit is proportional to the current flowing out of the supercapacitor to the load. Thus, by reading the analogue output from the differentiator, we can approximate the discharge current. The comparator circuit is used to ensure that the controller is notified only about activities that exceed a certain  threshold. For a constant power load, the current consumption from the capacitor will increase as the voltage decreases. Since the LMM tracks the current, it can act as a dynamic threshold detector capable of tracking the energy state of the buffer. 

For the circuit shown in Fig.~\ref{fig_load_monitor_circuit}, assuming $R_{F}  >> R_{1}$, output voltage $V_{DIFF}$ can be written as,

\begin{equation}
\label{eqn_differntiator_op}
    V_{DIFF} = -R_{F}C_{1}\frac{dv}{dt}
\end{equation}

The discharge current through the capacitor $C_{STORE}$  can be written as $I_{discharge} = C_{STORE} \frac{dv}{dt} $. Using Eq. \ref{eqn_differntiator_op}, we can write,

\begin{equation}
\label{eqn_discharge_current}
    I_{discharge} = C_{STORE}\frac{-V_{DIFF}}{R_{F}C{1}}
\end{equation}

Eq. \ref{eqn_discharge_current} implies that  by reading $V_{DIFF}$, the discharge current at any instance can be calculated. 

\subsection{InfiniteEn Control Unit}
The InfiniteEn Control Unit (ICU) is responsible for executing the state machines based on the inputs from  onboard modules. This includes selecting the charge-discharge routes of the supercapacitors,  handling interrupts from LMM and energy combiner, etc. We use a low-power ARM Cortex-M0+ microcontroller STM32L072 from STMicroelectronics \cite{stm32l072} to run the ICU state machine. STM32L072 is one of the low-power ARM processors available in the market. It provides multiple low-power modes like sleep, stop, standby, etc. along with a low power run mode in which the code can be executed at a reduced frequency to save energy. The ARM processor stays in sleep mode when the other modules are dormant and wakes up periodically or asynchronously to handle requests or inputs from the peripherals. 

\section{Evaluation}

\label{section_evaluation}

Evaluation of the InfiniteEn architecture involves the assessment of its individual modules for different test cases and scenarios. This section provides a detailed discussion on the evaluation of the InfiniteEn architecture.
\subsection{Energy Combiner}

The energy combiner is designed to \begin{inparaenum}[(i)]\item combine energy from multiple sources  \item estimate harvester current and \item measure the energy transferred from the source to the energy buffer\end{inparaenum}. We evaluated the efficiency of  the energy combiner for all three functionalities.  As discussed in Section \ref{section_infiniteEn_architecture}, the energy combiner uses two capacitors per source. We chose a pair of \SI{500}{\micro\farad} capacitors per source.  The capacitor cycles its charge-discharge between  \SI{2.5}{\volt} and \SI{3.2}{\volt}. We chose these values of thresholds to ensure that the TPS63900 regulator achieves 90\% or more efficiency. The threshold levels were chosen based on the efficiency curve from the datasheet and later confirmed  empirically. The datasheet claims an average efficiency of 90\% or more when the output current is between \SI{100}{\micro\ampere} and \SI{15}{\milli\ampere}.  A  Joulescope  was used to measure the energy transferred by the capacitor, $E_{cap}$ during a discharge and was measured as $E_{cap}= \SI{930}{\micro\joule}$. The maximum input current ($I_{limit}$) into the regulator was limited to \SI{5}{\milli\ampere}  using the input current limiter of TPS63900. The harvester sources were emulated using Keysight's DC power analyzer N6705B  in constant current mode~\cite{n6705b}.


\begin{figure}
\centering
    \subfloat[]{%
    \includegraphics[width=1.5in]{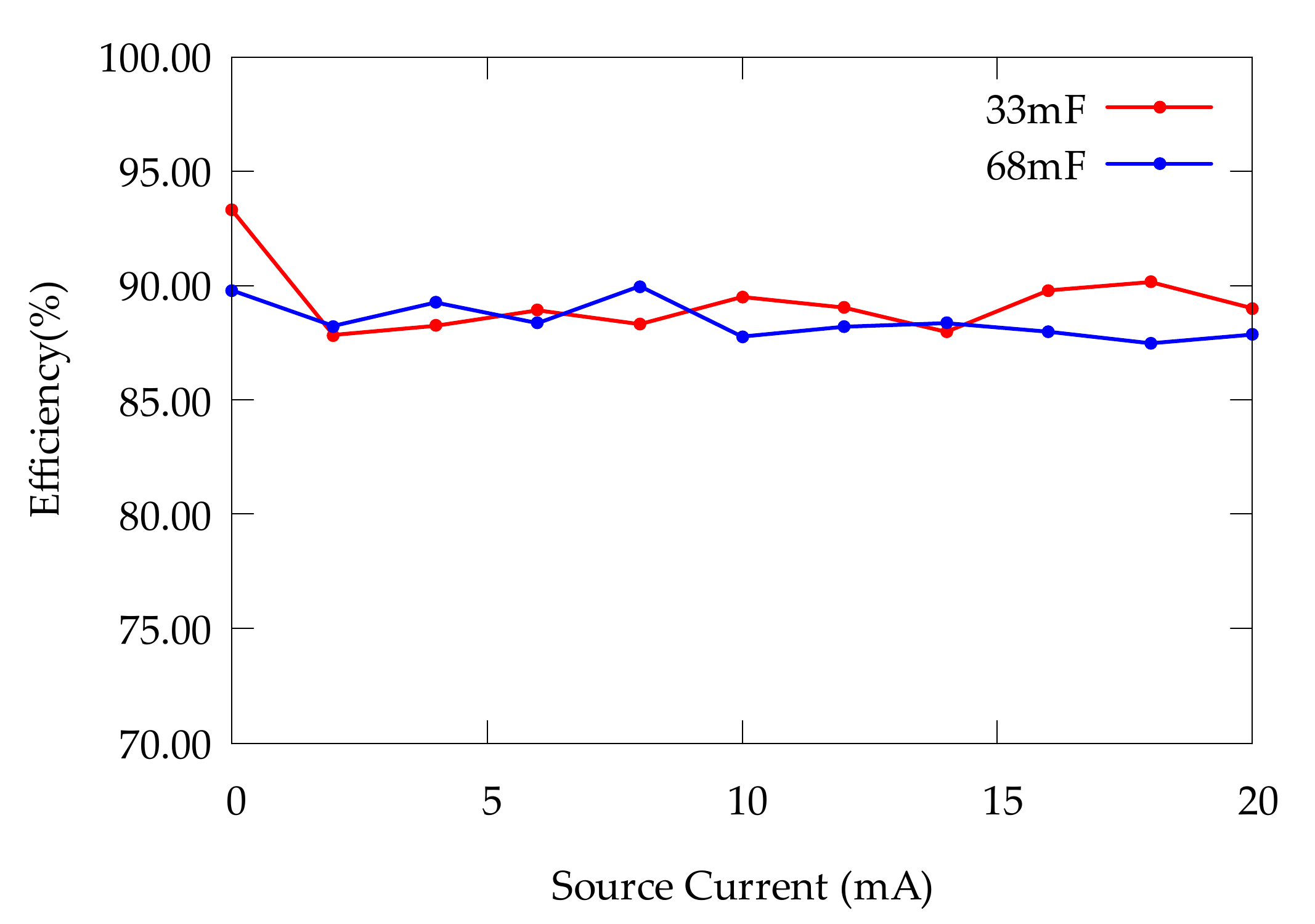}
    \label{fig_ec_efficiency}}
    \quad
    \subfloat[]{%
    \includegraphics[width=1.5in]{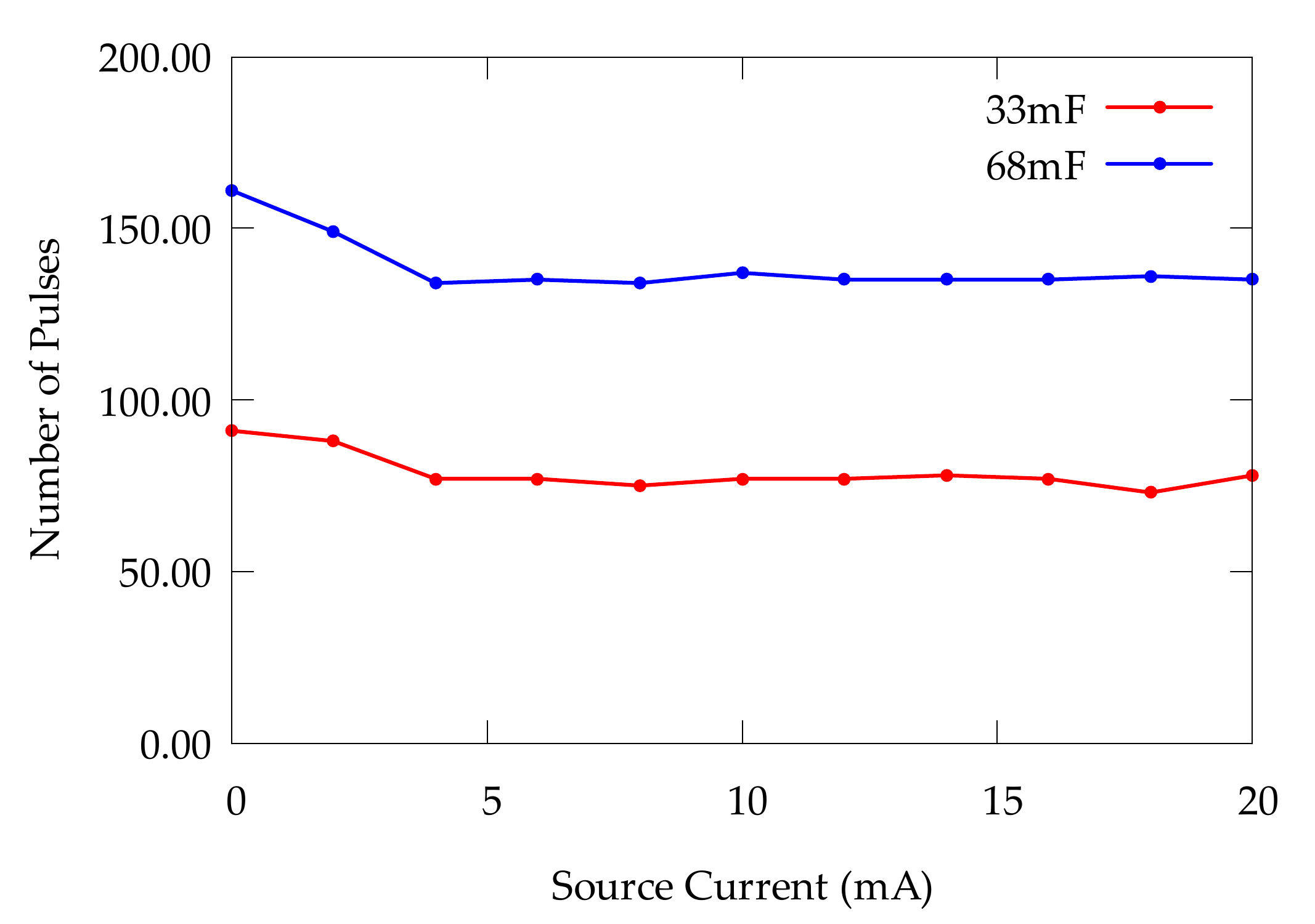}
    \label{fig_ec_current_vs_n}}
    \quad
    \subfloat[]{%
    \includegraphics[width=1.5in]{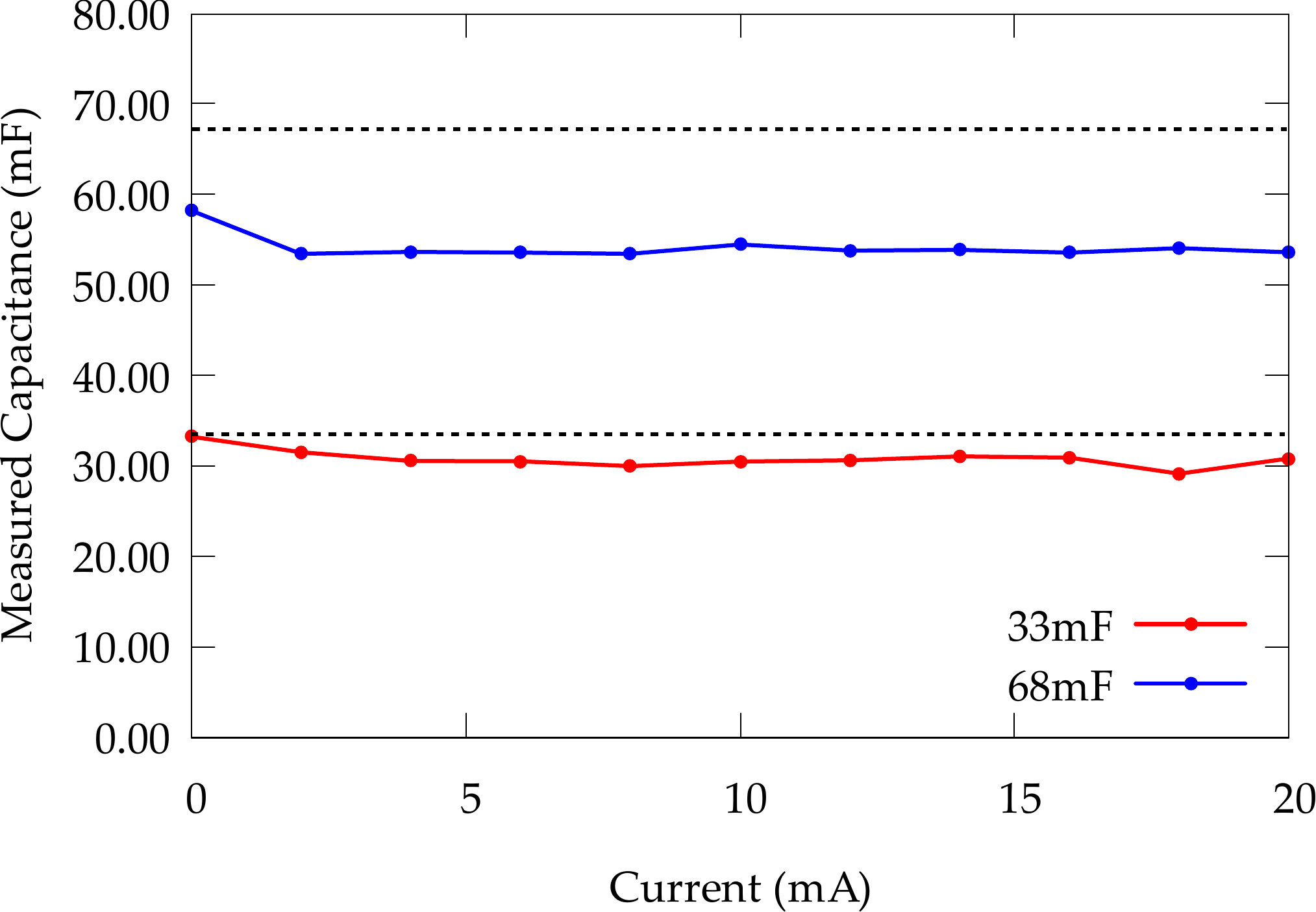}
    \label{fig_ec_measured_capacitance}}
    
    \caption{ Results from the experiments with energy combiner using temporary buffers of \SI{500}{\micro\farad} capacitors to combine two even sources and charging \SI{33}{\milli\farad} and \SI{68}{\milli\farad} supercapacitors. Energy harvester currents vary from \SI{0.005}{\milli\ampere} to \SI{20}{\milli\ampere} while the input current to switching regulator limited to \SI{5}{\milli\ampere}. \protect\subref{fig_ec_efficiency} efficiency of energy transfer  \protect\subref{fig_ec_current_vs_n} number of pulses received from the energy combiner \protect\subref{fig_ec_measured_capacitance} capacitance estimated based on the energy transferred to the capacitors; dashed lines representing manufacturer's values.}
    \label{ec_results}
    \vspace{-5mm}
\end{figure}

\textit{Efficiency:} The efficiency of the energy combiner is calculated as the ratio of total energy transferred to the energy buffer to the total energy consumed from the source. The efficiency calculated is plotted in Fig.~\ref{fig_ec_efficiency}. We observe that the energy combiner is able to achieve more than 90\% efficiency when the harvesting rate is extremely low. The efficiency stays at 88\% or more at higher harvesting rates. 

The number of energy transfers from the temporary buffer to the main storage buffer is shown in Fig.~\ref{fig_ec_current_vs_n}. From the figure, it could be observed that the total number of energy transfer cycles saturates for harvesting currents greater than \SI{5}{\milli\ampere}. This is due to the input current limit of TPS63900. The input current to TPS63900 was limited to \SI{5}{\milli\ampere}. Thus, any increase in source current after \SI{5}{\milli\ampere} does not impact the discharging rate of the temporary buffers.

\begin{figure}
    \centering    
    \subfloat[]{\includegraphics[width=1.5in]{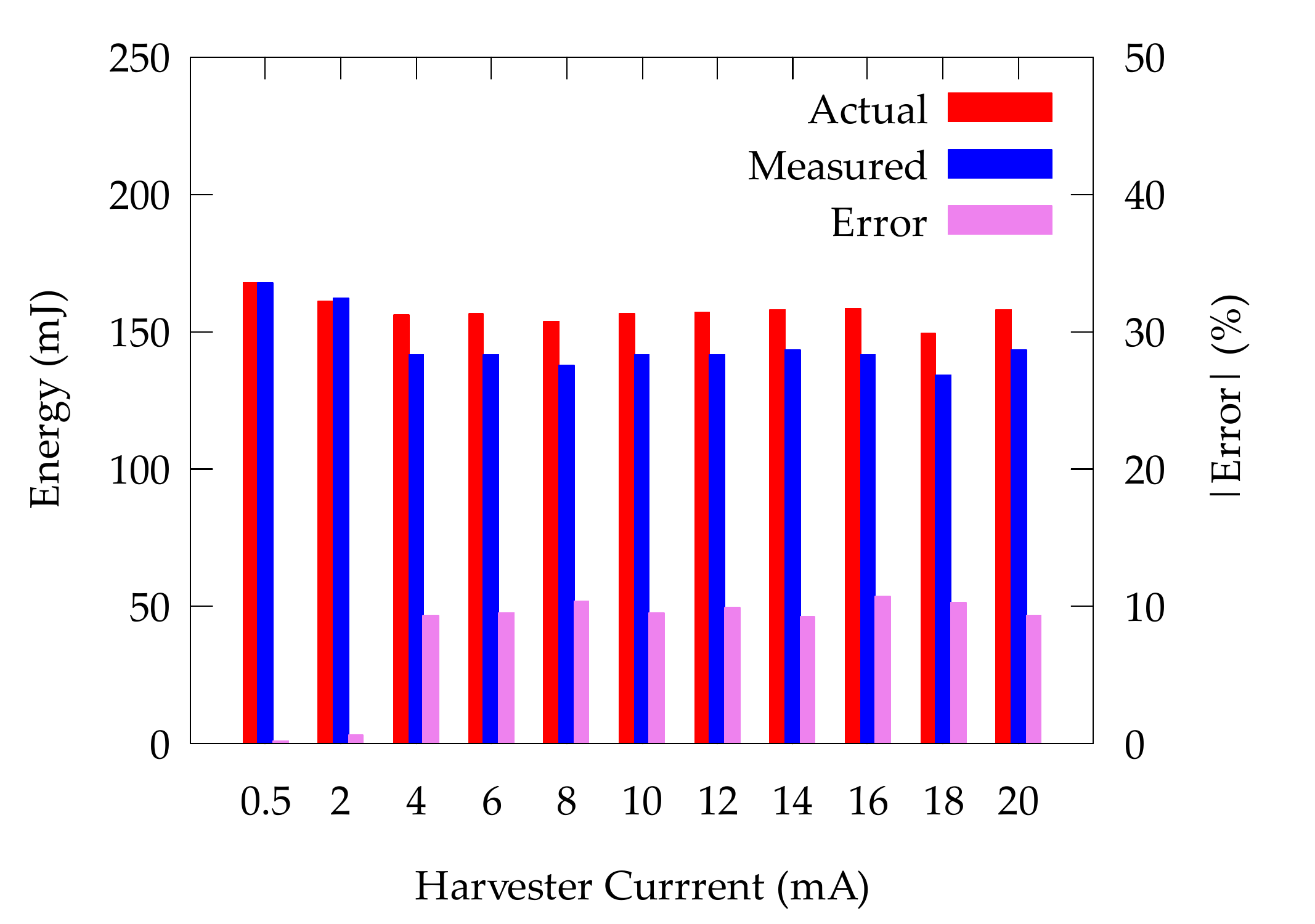}
    \label{fig_ec_33mF_energy_measurement}}    
   \quad     
     \subfloat[]{\includegraphics[width=1.5in]{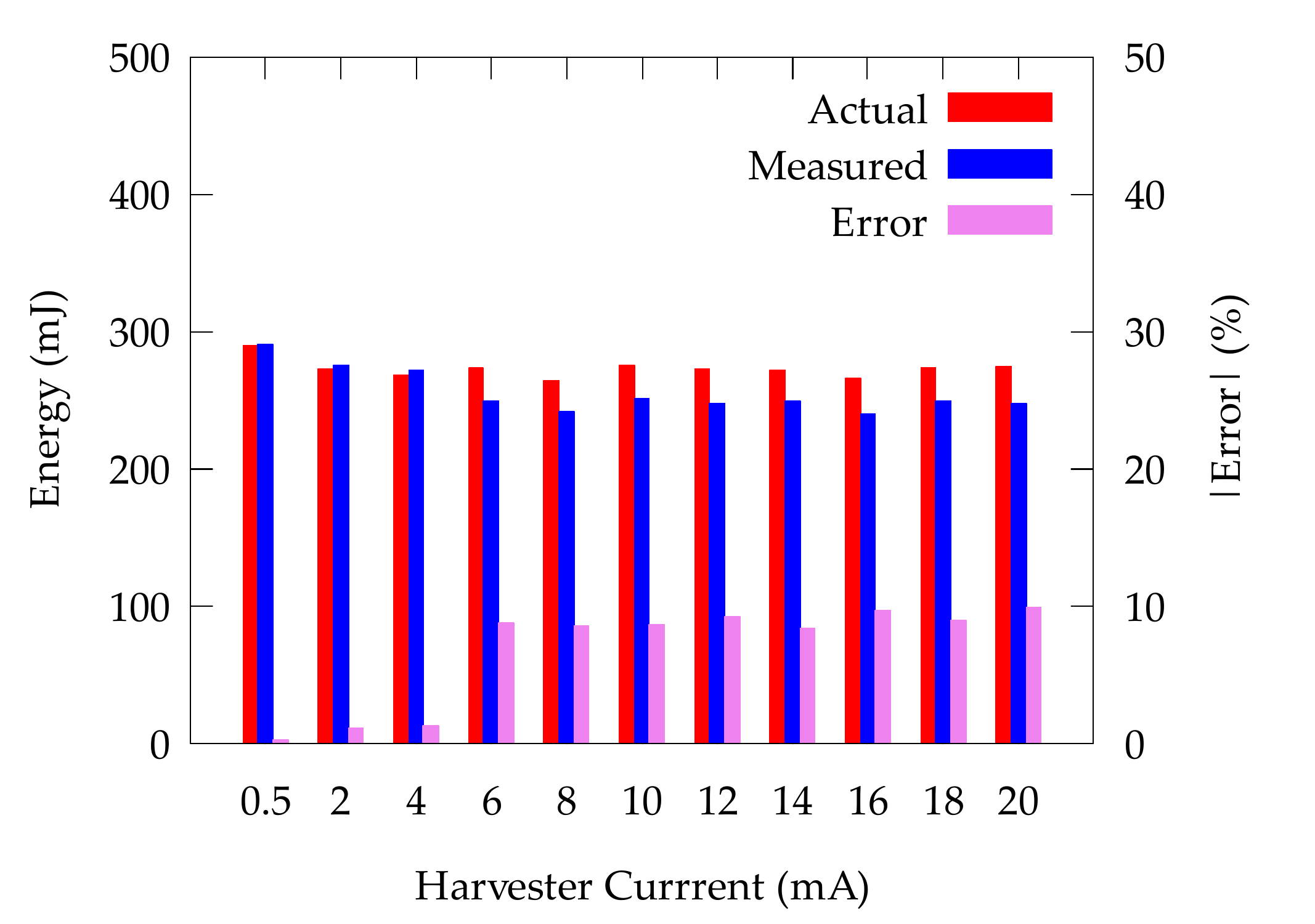}
    \label{fig_ec_68mF_energy_measurement}}

    \caption{Energy measured by the energy combiner with input current limit \SI{5}{\milli\ampere} and charging a \protect\subref{fig_ec_33mF_energy_measurement}
    \label{fig_ec_results_energy_measurement} \SI{33}{\milli\farad} capacitor \protect\subref{fig_ec_68mF_energy_measurement} \SI{68}{\milli\farad} Capacitor.}
    \vspace{-5mm}
\end{figure}

\begin{figure}
    \centering
    
    \subfloat[]{\includegraphics[width=1.5in]{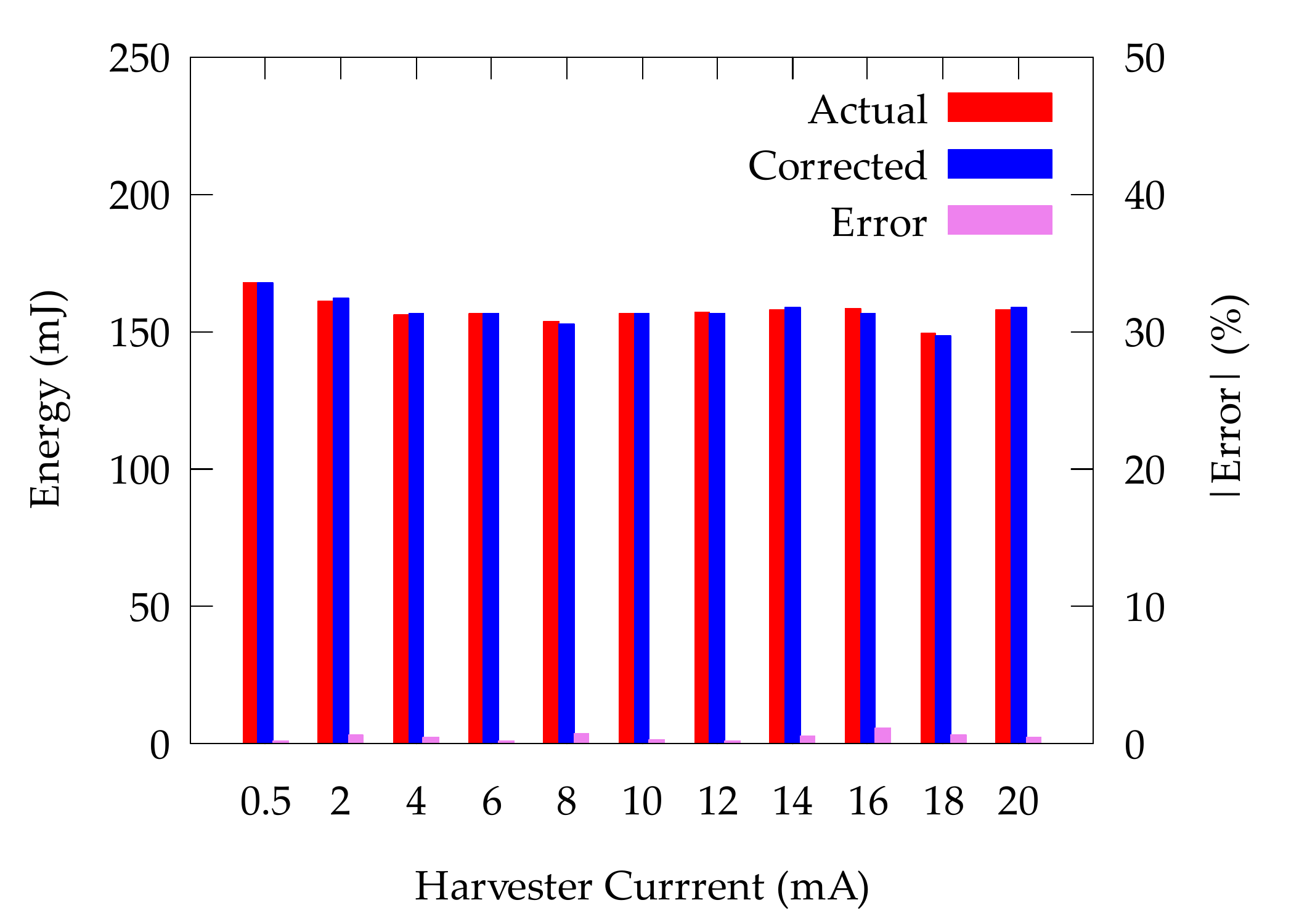}
    \label{fig_ec_33mF_energy_measurement_corrected}}    
   \quad     
     \subfloat[]{\includegraphics[width=1.5in]{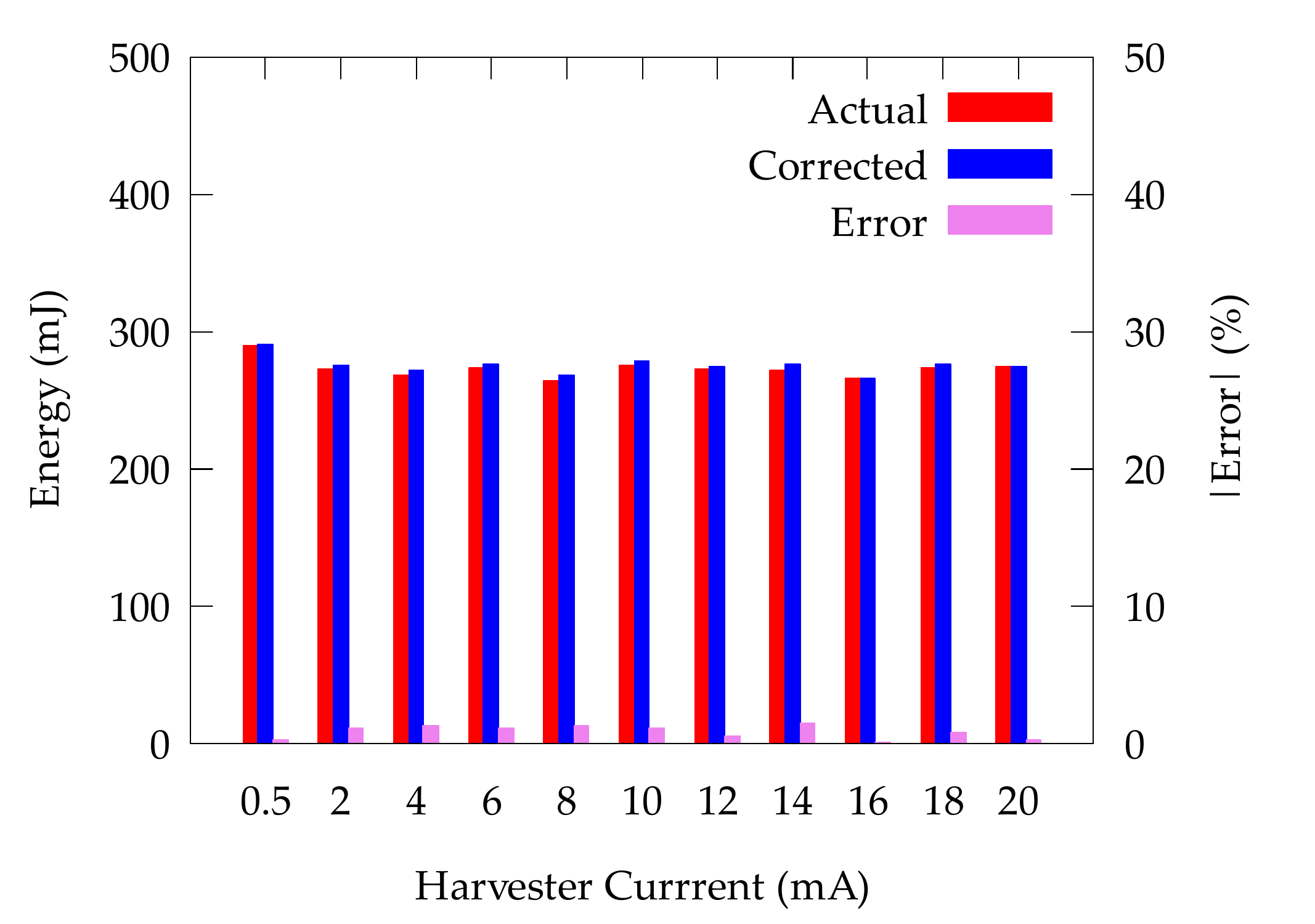}
    \label{fig_ec_68mF_energy_measurement_corrected}}

    \caption{Corrected energy measurement values with input current limit \SI{5}{\milli\ampere} and charging a \protect\subref{fig_ec_33mF_energy_measurement}
     \SI{33}{\milli\farad} capacitor \protect\subref{fig_ec_68mF_energy_measurement} \SI{68}{\milli\farad} Capacitor.}
    \label{fig_ec_results_energy_measurement_corrected}
    \vspace{-5mm}
\end{figure}

\textit{Energy Measurement:} Energy combiner estimates the amount of energy transferred to the storage buffer by counting the number of discharge cycles of the temporary buffers. Every discharge equals a fixed amount of energy transferred to the buffer.  As stated earlier, we used a temporary buffer of \SI{500}{\micro\farad} giving an energy transfer of $E_{cap} = \SI{930}{\micro\joule}$ per discharge.  The number of pulses from the energy combiner was recorded by the ICU, and the energy transferred was estimated by multiplying the number of pulses by $E_{cap}$.  The energy estimated during the experiment is shown in Fig.~\ref{fig_ec_results_energy_measurement}. The results show measurements close to the actual value for $I_{H} \leq I_{limit}$. For $I_{H} > I_{limit}$, more than 10\% error was noted. We investigated this issue and subsequently revealed that the energy transferred $E_{cap} = \SI{1030}{\micro\joule}$ for $I_{H} > I_{limit}$ and $E_{cap} = \SI{930}{\micro\joule}$ for $I_{H} \leq I_{limit}$. Upon further inspection, it was discovered that when $I_{H} > I_{limit}$, the discharging time of the temporary buffer exceeds the charging time. Thus, while one of the buffers is discharging, the other gets charged but keeps connected to the harvester until its pair is  discharged to $V_{L}$. As a result, the capacitor voltage reached slightly above the threshold \SI{3.20}{\volt}. It was measured to be \SI{3.28}{\volt}. So, we rectified the actual measurements by incorporating two conditional values, $E_{cap} = \SI{1030}{\micro\joule}$ for $I_{H} > I_{limit}$ and $E_{cap} = \SI{930}{\micro\joule}$ for $I_{H} \leq I_{limit}$. The corrected results are presented in Fig.~\ref{fig_ec_results_energy_measurement_corrected}.

\begin{figure}
    \centering
     \subfloat[]{%
     \includegraphics[width=1.5in]{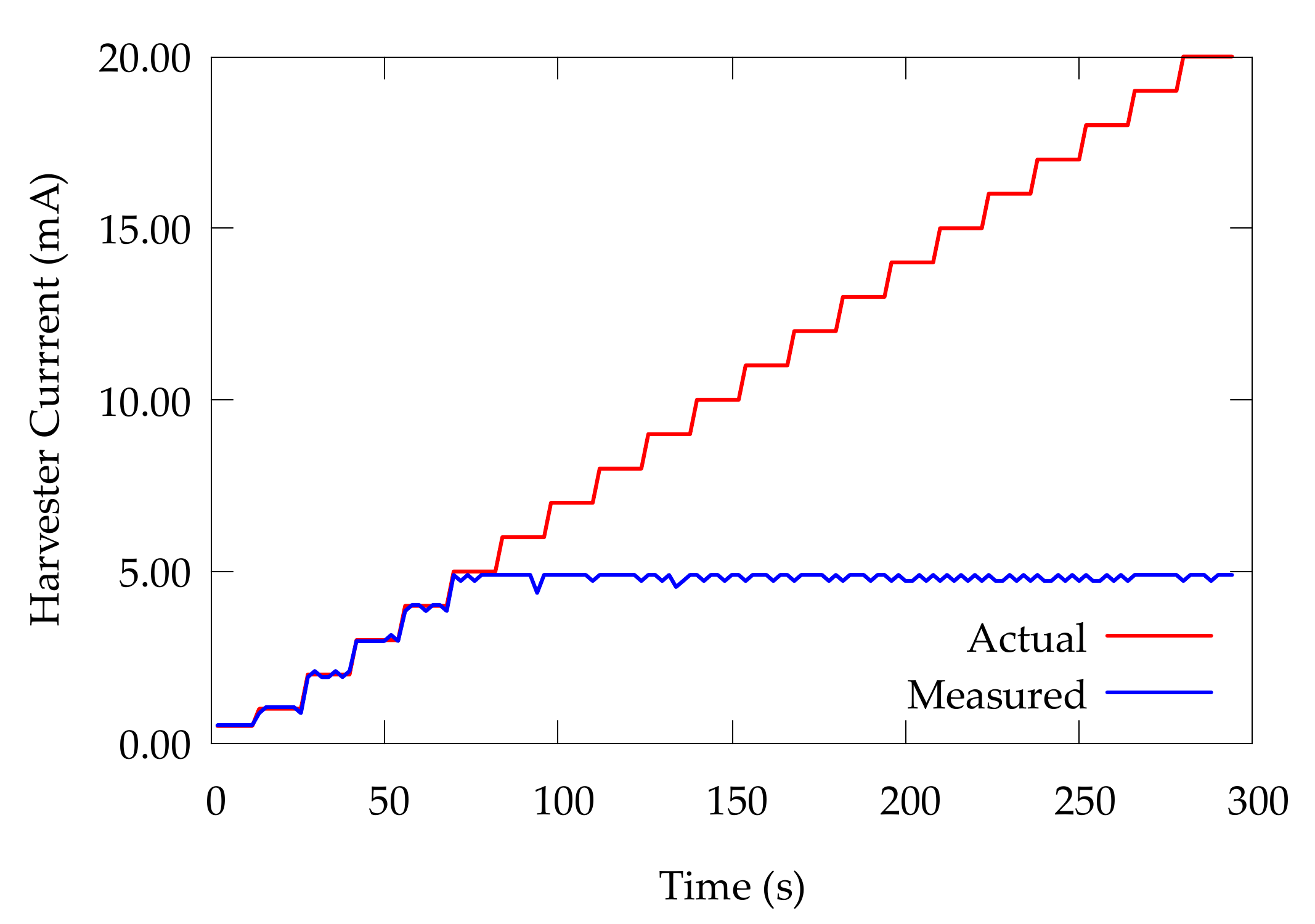}
    \label{fig_harvest_rate_5mA_limit}}    
    \quad 
     \subfloat[]{%
     \includegraphics[width=1.5in]{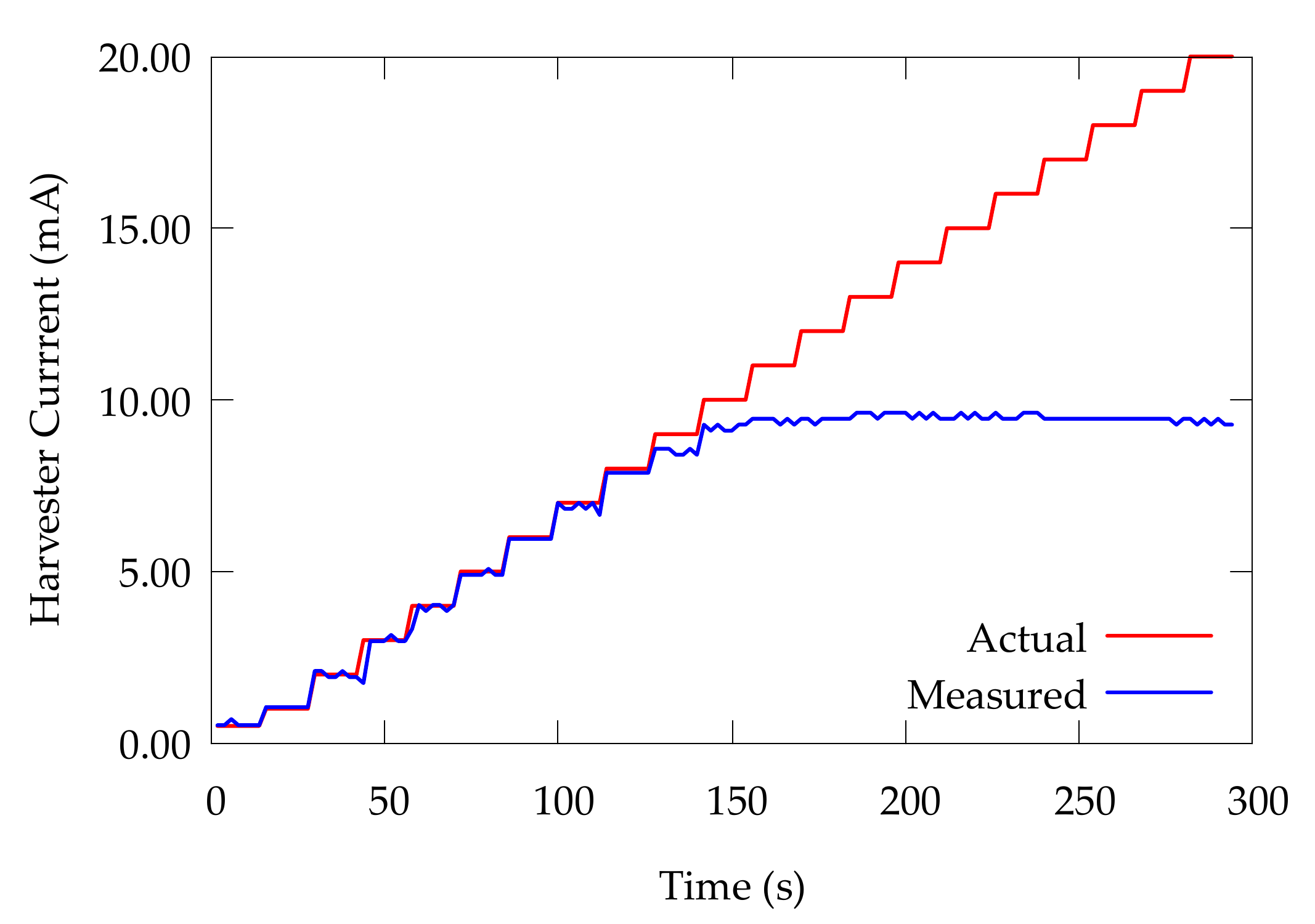}
    \label{fig_harvest_rate_10mA_limit}}
    \quad
    \caption{Harvesting current sensed by the energy combiner with input current limited to \protect\subref{fig_harvest_rate_5mA_limit} \SI{5}{\milli\ampere} \protect\subref{fig_harvest_rate_10mA_limit} \SI{10}{\milli\ampere} while varying combined source currents  from \SI{0.005}{\milli\ampere} to \SI{20}{\milli\ampere}.}
    \label{fig_ec_results_harvest_rate}
    \vspace{-5mm}
\end{figure}

\textit{Harvester Current Sensing:} To estimate the harvester current, Eq. \ref{eqn_average_harvester_current} can be used.  We conducted experiments to measure the harvester current by varying it from \SI{0.5}{\milli\ampere} to \SI{20}{\milli\ampere} in steps of \SI{1}{\milli\ampere} every 15 seconds.  The experiments were conducted with input current limit, $I_{limit}=\SI{5}{\milli\ampere}$ and $I_{limit}=\SI{10}{\milli\ampere}$. The results are plotted in Fig.~\ref{fig_ec_results_harvest_rate}. The estimates give the combined harvester current from both sources.

The results follow a similar pattern as in energy measurements. Estimated harvester current saturates for $I_{H} >= I_{limit}$. This is due to the input current limit of the switching regulator. The regulator trims down any input current higher than  $I_{limit}$. Consequently, the discharging time dominates the charging time and the effective input current to the energy combiner is, $I_{limit}$. As a result, in Fig.~\ref{fig_harvest_rate_5mA_limit}, for all $I_{H} >= \SI{5}{\milli\ampere}$, estimated current  $\approx \SI{5}{\milli\ampere}$. Similarly, for measurements with $I_{limit}=\SI{10}{\milli\ampere}$, estimated current  $\approx \SI{10}{\milli\ampere}$ for  $I_{H} >= \SI{10}{\milli\ampere}$ as shown in Fig.~\ref{fig_harvest_rate_10mA_limit}.

\subsection{Load Monitoring Module}

InfiniteEn uses a re-configurable architecture for the storage buffer. This gives the system the liberty to choose a storage buffer based on the incoming energy and the load demands. The reconfiguration, however, should happen as fast as possible to ensure uninterrupted energy flow to the load. Once the LMM detects a discharge, it triggers an interrupt and based on the discharge current, the ICU selects a new storage buffer. Thus, the reconfiguration time is mainly determined by the responsiveness of the LMM to discharge current and the time taken by the software task to perform the switchover from one buffer to the other. We conducted multiple experiments to evaluate the   LMM for its responsiveness, minimum detectable discharge and effectiveness in identifying load abnormalities. 

\begin{figure}
\centering
    \subfloat[]{ \includegraphics[width=1.5in]{ 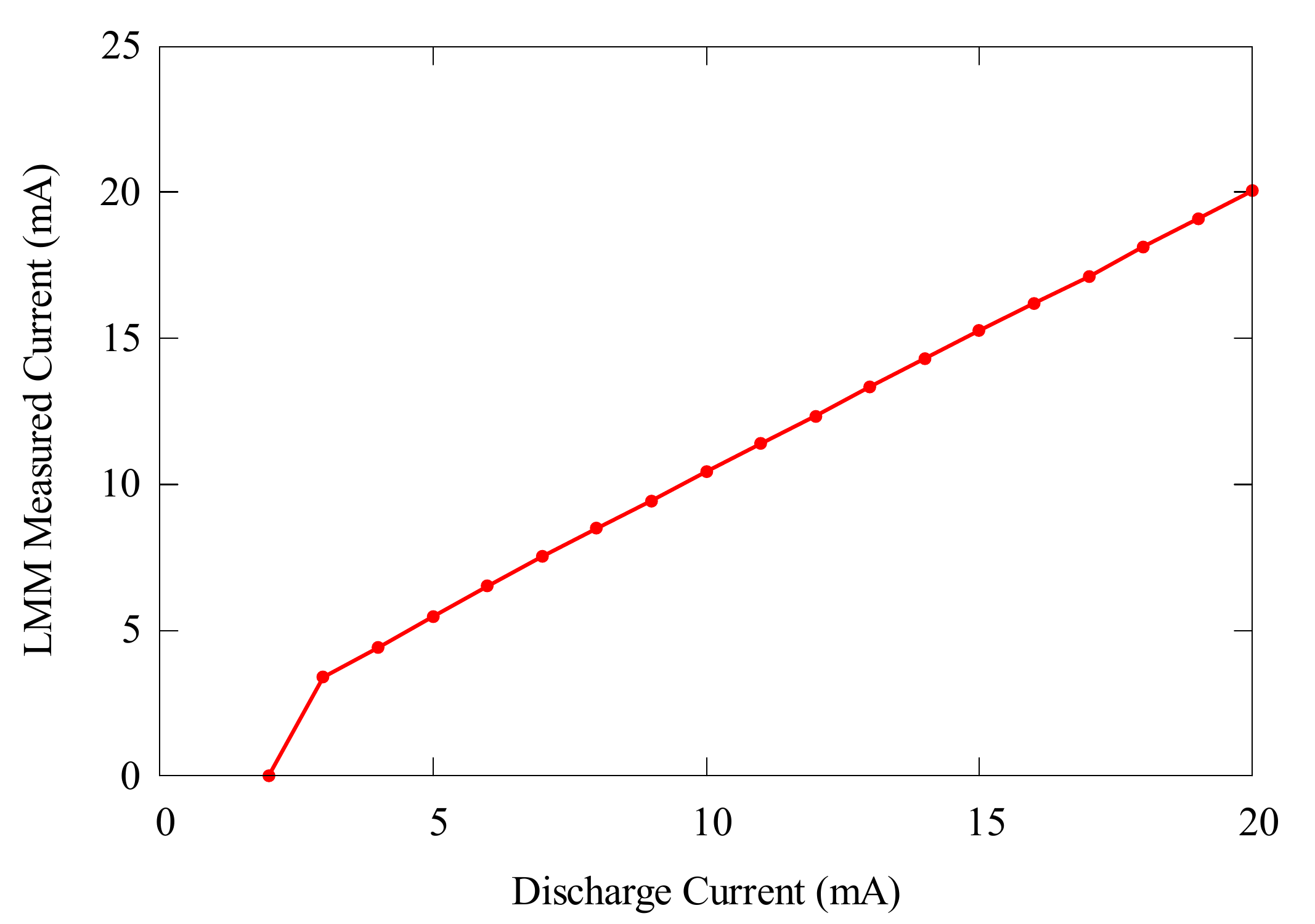}}
    \subfloat[]{ \includegraphics[width=1.5in]{ 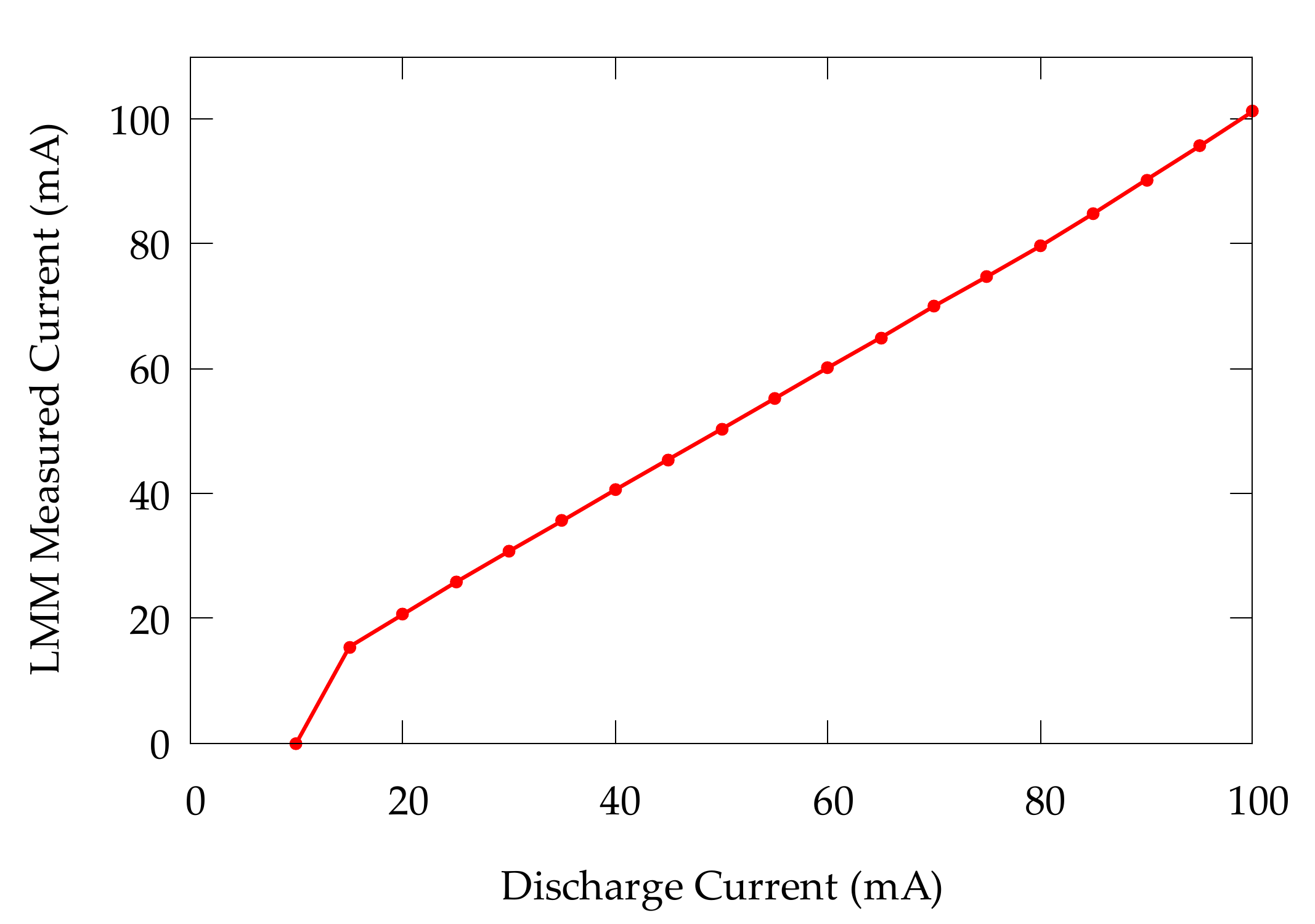}}
    
    \caption{Discharge current measured by the load monitoring module for discharge currents varying from (a) \SI{2}{\milli\ampere} to \SI{20}{\milli\ampere} for a \SI{5}{\milli\ampere} capacitor and (b) \SI{5}{\milli\ampere} to \SI{105}{\milli\ampere} for a \SI{100}{\milli\farad} capacitor.}
    \label{fig_lmm_measured_vs_actual}
    \vspace{-5mm}
\end{figure}

In the first experiment, the storage capacitor $C_{STORE}$ was charged to \SI{3}{\volt} and then  discharged with current pulses of \SI{30}{\milli\second} duration. We used two supercapacitors, \SI{15}{\milli\farad} and \SI{100}{\milli\farad}. Discharge current pulses varying from \SI{2}{\milli\ampere} to \SI{20}{\milli\ampere} were applied to  \SI{15}{\milli\farad} capacitor, and pulses varying from \SI{5}{\milli\ampere} to \SI{105}{\milli\ampere} were applied to  \SI{100}{\milli\farad} capacitor. A \SI{500}{\milli\second} silent period was kept between each pulse. The discharge current, capacitor voltage and LMM response were recorded using the Joulescope. Further, the analogue output from the LMM was used to calculate the discharge current. A plot of the actual discharge current vs the measured discharge current is shown in Fig.~\ref{fig_lmm_measured_vs_actual}.

\textit{Minimum detectable discharge:}
The results presented in  Fig.~\ref{fig_lmm_measured_vs_actual} reveal that there is a minimum discharge from the capacitor, below which the LMM does not respond. The \SI{15}{\milli\farad} supercapacitor must be discharged at least with a \SI{3}{\milli\ampere} current pulse of  \SI{30}{\milli\second} duration.  For a\SI{100}{\milli\farad} capacitor as shown in Fig.~\ref{fig_lmm_measured_vs_actual}, the discharge current should be  \SI{15}{\milli\ampere} and should last for \SI{30}{\milli\second}. The minimum detectable change of the LMM is determined by \begin{inparaenum}[(i)] \item the reference voltage of the comparator,  \item the values of $R_{F}$ and $C{1}$ or the gain of the differentiator  and \item  the capacity of the supercapacitor \end{inparaenum}. The right combinations of these components can be selected based on the maximum expected load discharge and the size of supercapacitor.


It must also be noted that the LMM, at this stage, can give only an  approximation of the discharge current to assist the control unit in managing the storage buffers. The current measurement accuracy of LMM is impacted at low discharge rates due to the noise, and we observe a maximum error of 12\% when measuring a \SI{3}{\milli\ampere} discharge current.

\begin{figure}
    \centering 
    \subfloat[]{\includegraphics[width=1.5in]{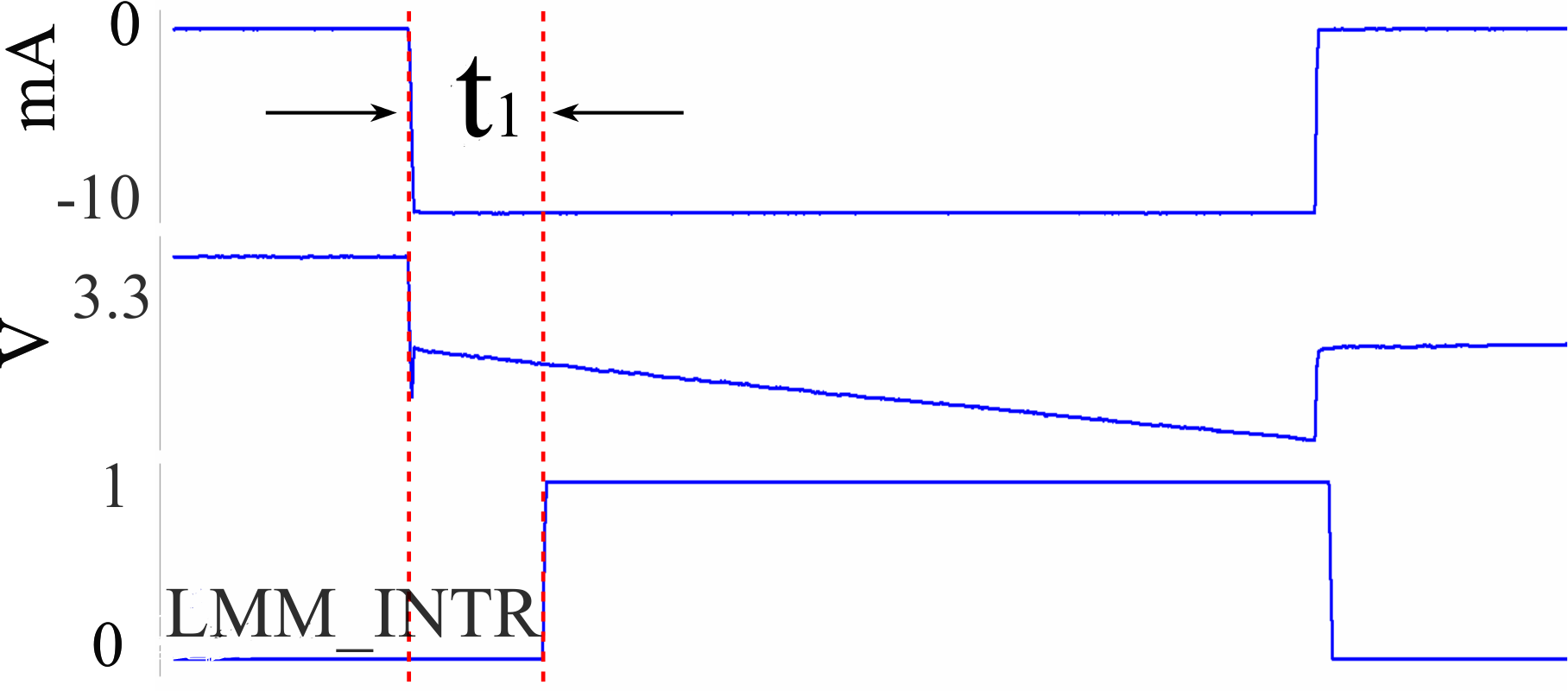}
    \label{fig_lmm_intr_closeup_15mA}}
    \quad
    \subfloat[]{\includegraphics[width=1.5in]{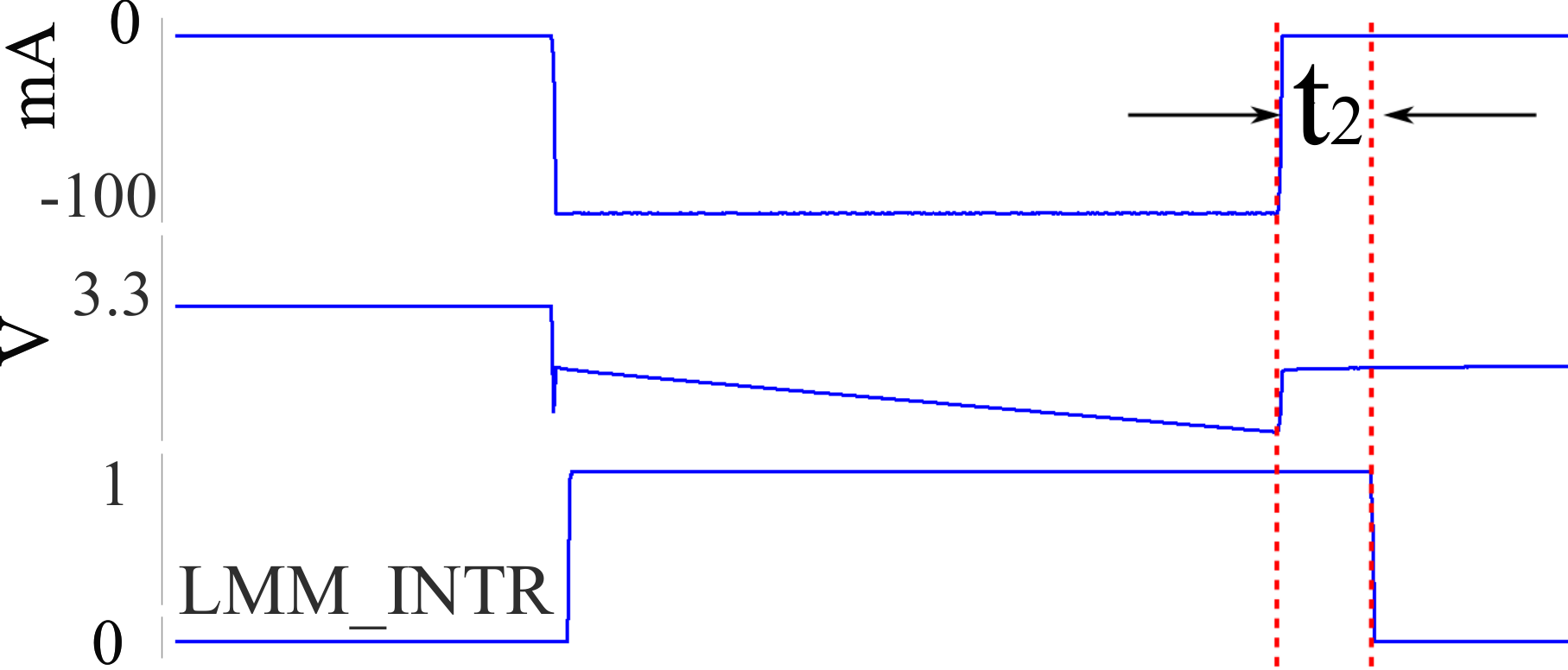}
    \label{fig_lmm_intr_closeup_100mA}}
    \caption{The response of the LMM when discharging a \SI{100}{\milli\farad} capacitor with \protect\subref{fig_lmm_intr_closeup_15mA}\SI{15}{\milli\ampere} current \protect\subref{fig_lmm_intr_closeup_100mA} \SI{100}{\milli\ampere} current pulses. }
    \vspace{-5mm}
\end{figure}

\textit{Response time:} We take a detailed look at the responses of the LMM to a low \SI{15}{\milli\ampere} current pulse in  Fig.~\ref{fig_lmm_intr_closeup_15mA} and  to a \SI{100}{\milli\ampere} current pulse in Fig.~\ref{fig_lmm_intr_closeup_100mA}. In Fig.~\ref{fig_lmm_intr_closeup_15mA}, $t_{1}$ is the time taken by LMM to respond to a slowly changing input corresponding to a \SI{15}{\milli\ampere} current pules. Similarly, in \ref{fig_lmm_intr_closeup_100mA}, LMM takes $t_{2}$ seconds to return to its default state after $t_{2}$ seconds. But, the response to the input change appears immediately at the output. Hence, it is evident that the LMM does not immediately respond to a slowly changing input, but with some  delay. On the other hand, a rapid change in the input is detected immediately, but it takes more time to go back to the default state after the input settles down. Delays $t_{1}$ and $t_{2}$ are due to the time taken by the input capacitor to charge and discharge.  From the experiments, the maximum value of $t_{1}$ was measured as \SI{10.32}{\milli\second} and $t_{2}$ was measured as \SI{7.63}{\milli\second}. It is possible to reduce both delays by optimizing the values of $R_{F}$ and $C_{1}$ or reducing the reference $V_{REF}$ of the comparator. A smaller value of $R_{F}$ and/or $C_{1}$ will reduce the gain and consequently the time $t_{2}$ by reducing the time taken by the capacitor $C_{1}$ to discharge. Similarly, a smaller value of $V_{REF}$ will reduce $t_{1}$.

\begin{figure*}
    \centering 
    \subfloat[]{    \includegraphics[width=3.0in]{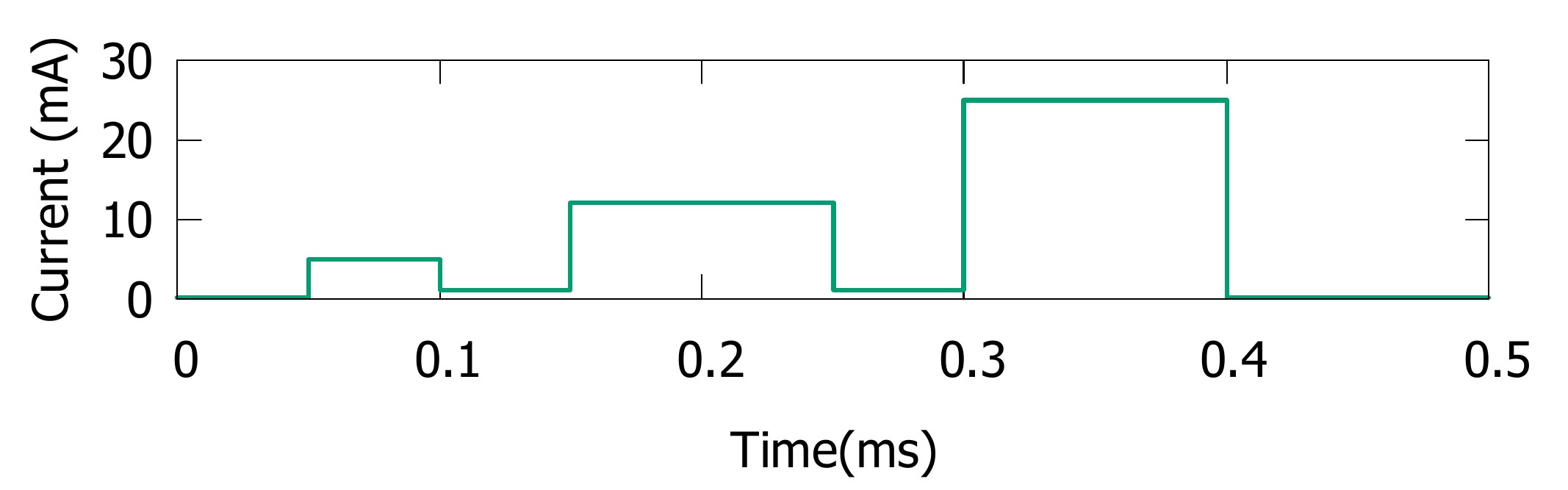}
    \label{fig_test_profile_regular}}
    \subfloat[]{    \includegraphics[width=3.0in]{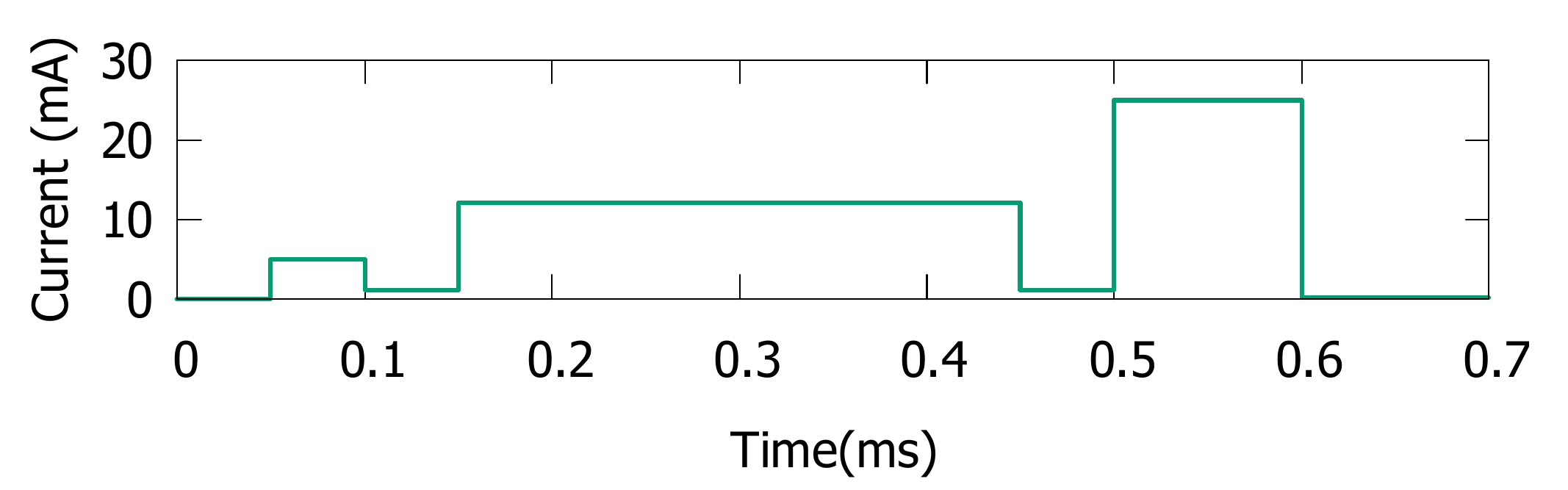}
    \label{test_profile_with_bug}}

    \caption{Energy profile used for assessing the effectiveness of LMM in detecting load abnormalities \protect\subref{fig_test_profile_regular} a regular profile with no load abnormalities \protect\subref{test_profile_with_bug}  a profile where one of the tasks executes for a prolonged period due to a software or hardware issue.}
    \label{fig_test_profile}
    \vspace{-5mm}
\end{figure*}

\begin{figure*}
\centering

    \subfloat[]{\includegraphics[width=3.0in]{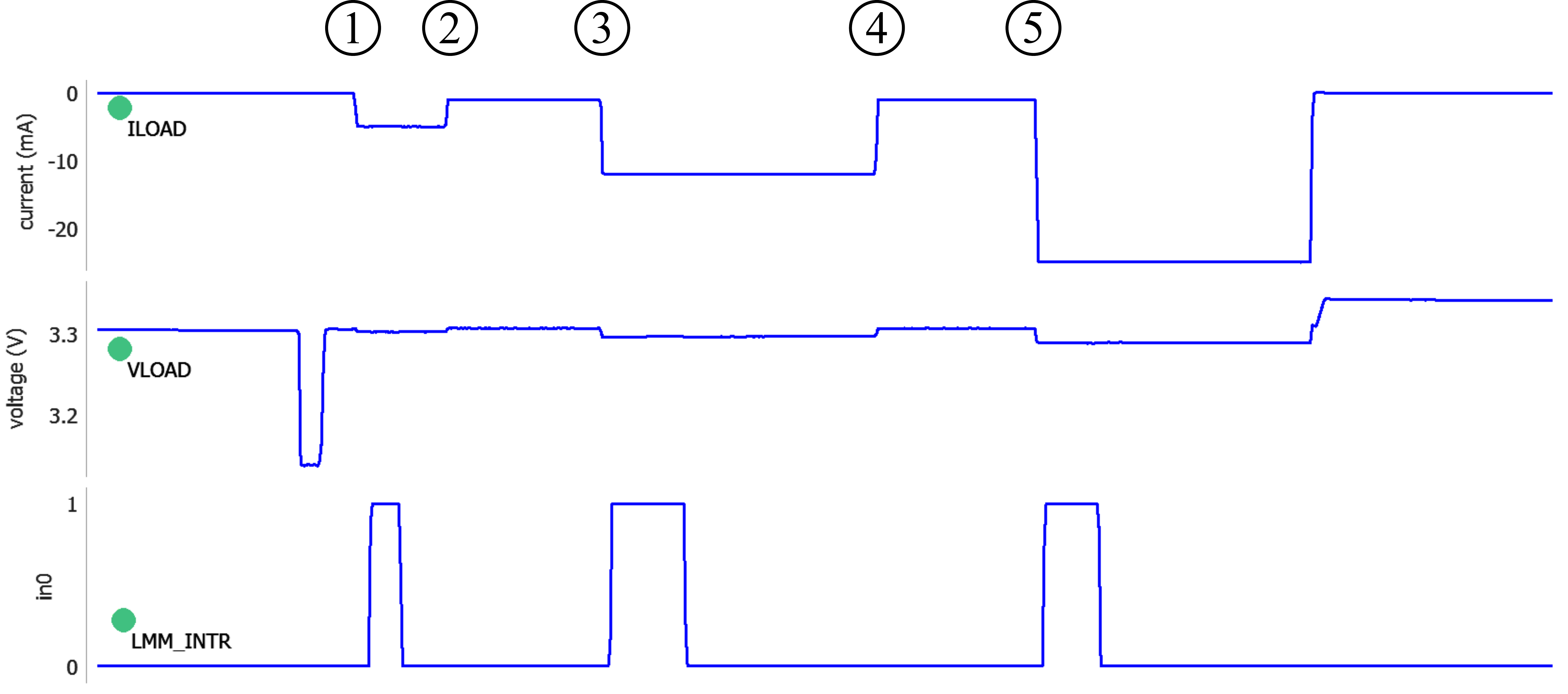}
    \label{fig_lmm_no_bug_js}}
    \quad	
    \subfloat[]{\includegraphics[width=3.0in]{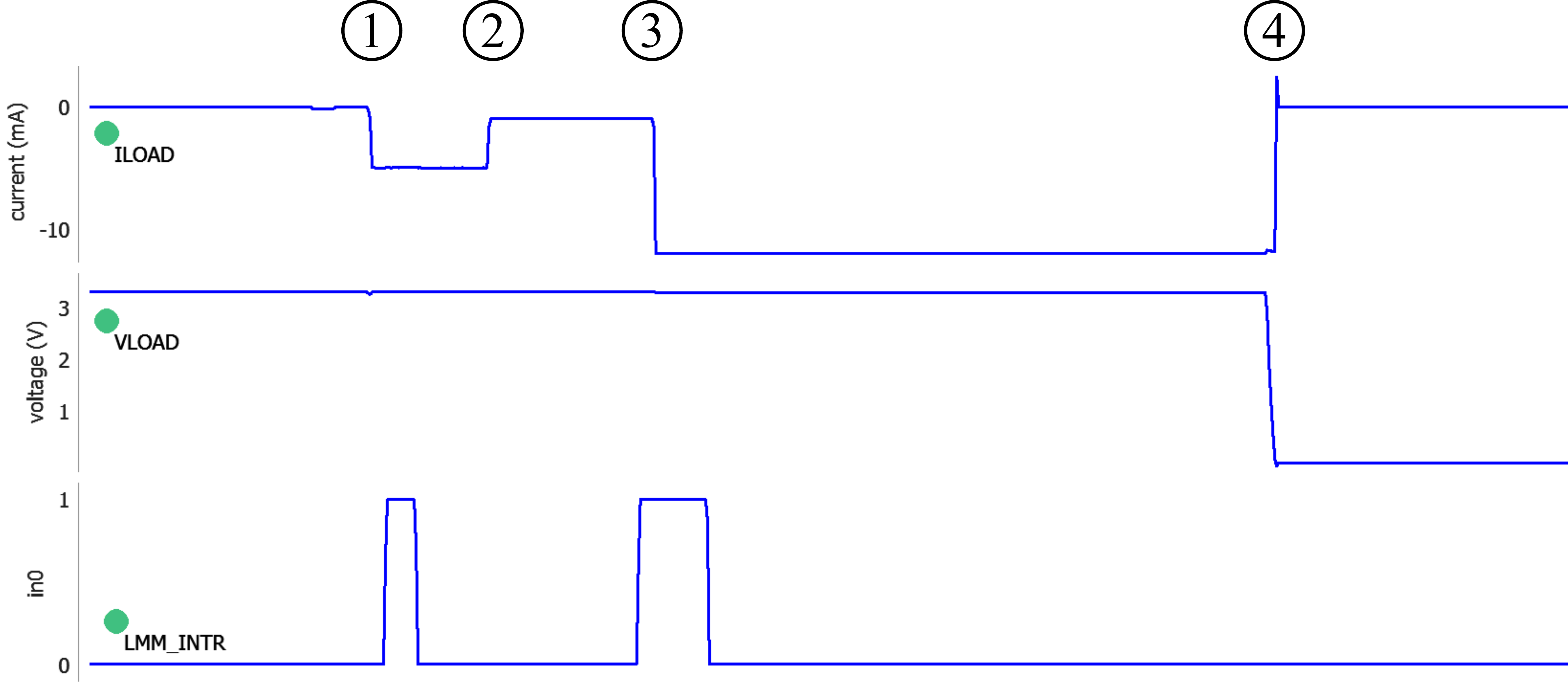}
    \label{fig_lmm_bug_js}}   
    
    \caption{Response of  LMM recorded using Joulescope when \protect\subref{fig_lmm_no_bug_js}   the regular profile is running  \protect\subref{fig_lmm_bug_js} defective profile is running.}

    \label{fig_lmm_response_to_test_profile_js}
    \vspace{-5mm}
\end{figure*}

\begin{figure*}
\centering

    \subfloat[]{\includegraphics[width=3.0in]{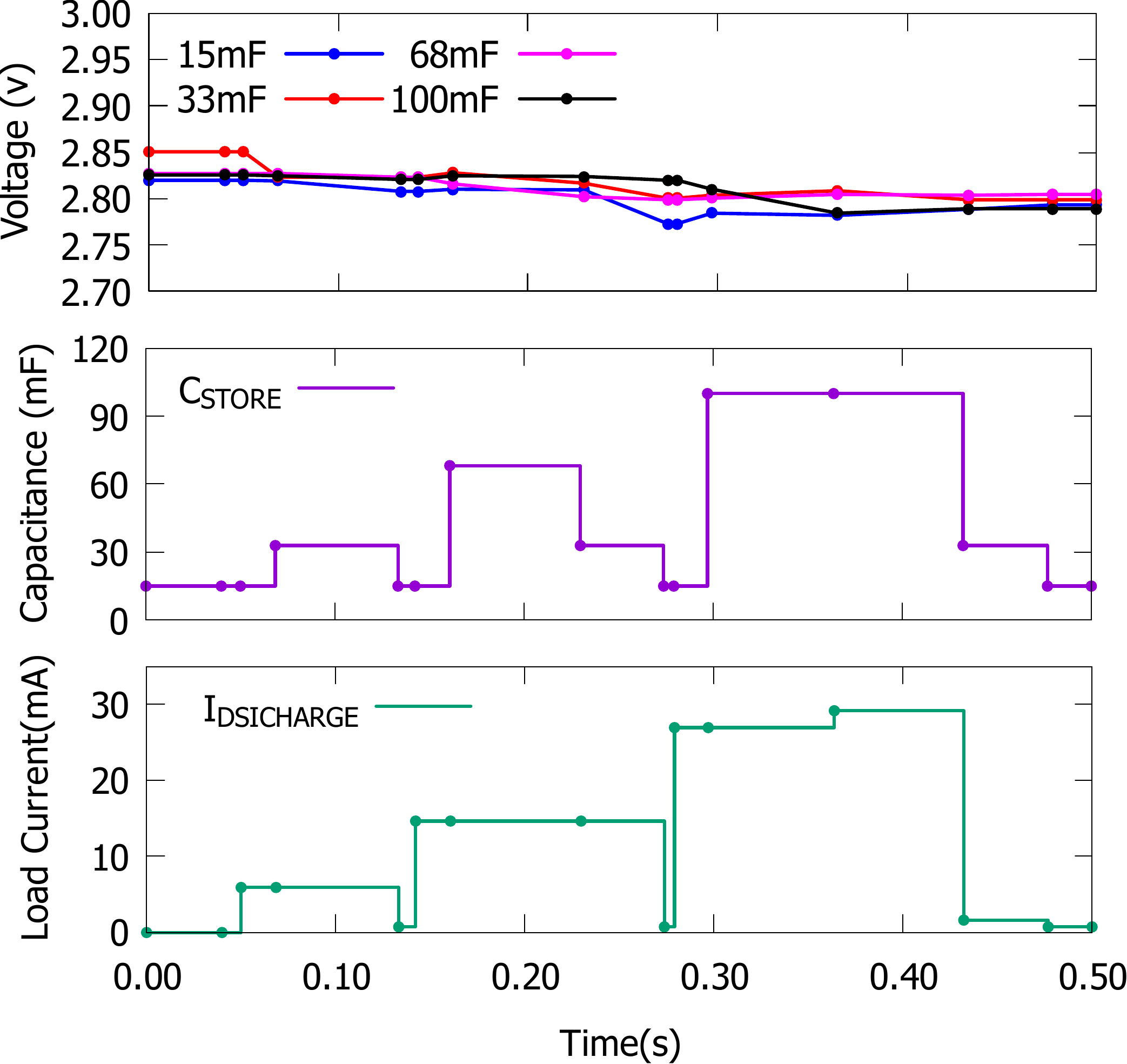}
    \label{fig_lmm_no_bug_icu}}
    \quad	
    \subfloat[]{\includegraphics[width=3.0in]{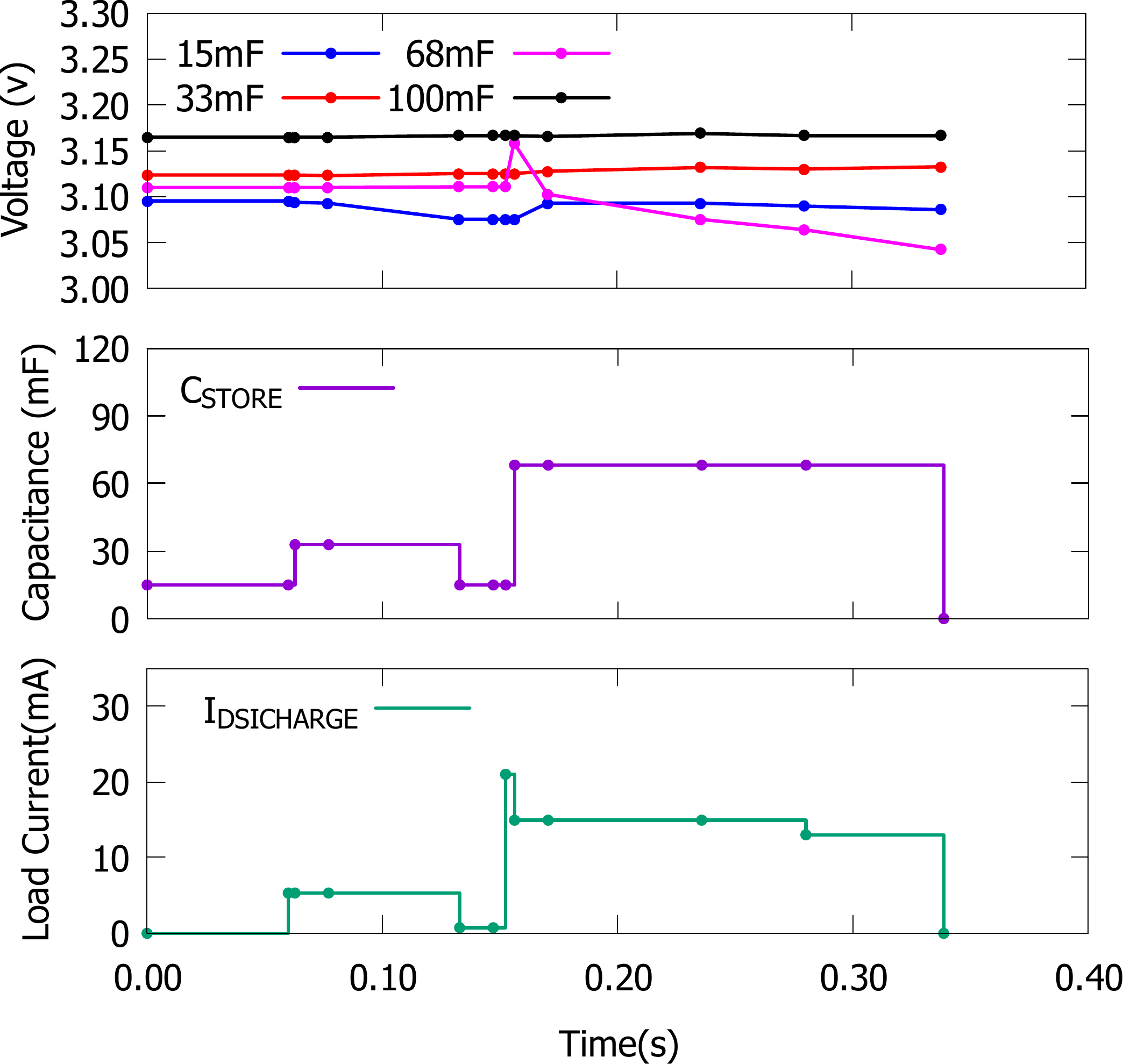}
    \label{fig_lmm_bug_icu}}   
    
    \caption{Snapshot of capacitor array voltages,  discharge current and configured capacitance of InfiniteEn when \protect\subref{fig_lmm_no_bug_icu}  the regular profile is running \protect\subref{fig_lmm_bug_icu} defective profile is running. }

    \label{fig_lmm_response_to_test_profile_icu}
        
\end{figure*}

\textit{Load Abnormality Detection:} LMM detects load abnormalities by tracking the energy states of the load. To evaluate LMM's efficiency in abnormality detection, we simulated the energy profile of a load using the N6705B Power Analyzer. We defined two profiles, one for regular execution and the other for a buggy execution, as shown in Fig.~\ref{fig_test_profile}. The regular profile defines 4 tasks \begin{inparaenum}[(i)] \item $task1$ which consumes \SI{5}{\milli\ampere} and runs for \SI{50}{\milli\second} \item $task2$ which consumes \SI{1}{\milli\ampere} and runs for \SI{50}{\milli\second} andgets executed twice \item $task3$ which consumes \SI{12}{\milli\ampere} and runs for \SI{100}{\milli\second}  and \item $task4$ which consumes \SI{25}{\milli\ampere} and runs for \SI{100}{\milli\second} \end{inparaenum}. For the corrupted execution, the execution of $task3$  is extended by \SI{200}{\milli\second}. The Joulescope was used to record the load voltage, actual load current and the interrupts from the LMM. In addition, the ICU measured the voltage of the supercapacitors, the discharge current using LMM and the total capacitance available to the load. The ICU was configured to measure discharge current in every \SI{50}{\ms} after a task was started. Switching between capacitors was done based on a look-up table,  as shown in Table \ref{tab_cstore_lookup}. The lookup table defines the maximum current allowed for each capacitor in  the array. The experiments were conducted for both test profiles. Details collected using the Joulescope are depicted in Fig.~ \ref{fig_lmm_response_to_test_profile_js} and that reported by the ICU are shown in Fig.~\ref{fig_lmm_response_to_test_profile_icu}. The graphs plotted using Joulescope data in Fig.~ \ref{fig_lmm_response_to_test_profile_js} show the actual energy consumption pattern of the target and the plots in Fig.~\ref{fig_lmm_response_to_test_profile_icu} show the pattern tracked by the LMM.

\begin{table}[]
\caption{Look up table used for $C_{STORE}$ selection\label{tab_cstore_lookup}}
\begin{center}
\begin{tabular}{ |c|c|c| } 
 \hline
 Capacitor  & Capacitance (mF) & Current range (mA) \\ 
 \hline
 $C_{1}$ & 15 &  $i \leq \SI{1}{\milli\ampere}$ \\ 
 $C_{2}$ & 33 & $ \SI{1}{\milli\ampere} < i   \leq \SI{10}{\milli\ampere}$    \\ 
 $C_{3}$ & 68 & $ \SI{10}{\milli\ampere} < i   \leq \SI{20}{\milli\ampere}$\\
 $C_{4}$ & 100 & $ i >  \SI{20}{\milli\ampere}$\\
 \hline
\end{tabular}
\end{center}
\vspace{-4mm}
\end{table}

In Fig.~\ref{fig_lmm_no_bug_js} and Fig.~\ref{fig_lmm_bug_js}, the time stamps corresponding to each task  are marked from $1-5$. Corresponding measurements from the ICU are shown in Figs.~\ref{fig_lmm_no_bug_icu} and Fig.~\ref{fig_lmm_bug_icu}. Initially, the target is in sleep mode and consumes no or minimum energy. The four capacitor voltages were measured to be  $V_{C1} = \SI{2.82}{\volt}, V_{C2}  = \SI{2.85}{\volt}, V_{C3}  = \SI{2.83}{\volt}, V_{C4}  = \SI{2.89}{\volt}$ at the beginning. The target was supplied from the \SI{15}{\milli\farad} capacitor ($C_{1}$), ie. $C_{STORE} = \SI{15}{\milli\farad}$. 

In the first experiment, $task1$ of the regular profile was initiated at instance $1$ as shown in Fig.~\ref{fig_lmm_no_bug_js}. In response to this, the  LMM generates an interrupt  which results in changing $C_{STORE}$ to $C_{2}$. This also brings down the LMM interrupt since the rate of change in voltage is now below the threshold levels. The discharge currents were recorded for both the rising edge and falling edge of the LMM interrupt. The ICU, before going back to sleep, starts a timer to measure the discharge current in every \SI{50}{\milli\second}. At the instance, $2$, $task2$ was started but did not get detected by the LMM since the current consumption was less than the minimum detectable value for $C_{2}$. Looking from Fig.~\ref{fig_lmm_no_bug_icu}, it is clear that no $C_{STORE}$ adjustment  happened at instance $2$. Switching $C_{STORE}$ happened only after the previously started timer was fired, and current consumption was measured. Thus, the discharge current measurements by LMM show a longer period for $task1$ and a shorter period for $task2$, which is depicted in Fig.~\ref{fig_lmm_no_bug_icu}. Similarly, execution  of $task3$ at instance $3$ and $task4$ at instance $5$ were detected by the LMM and $C_{STORE}$ was adjusted accordingly. For $task3$, $C_{3}$ wass used and for $task4$, $C_{4}$ was used. Again, instance $4$ corresponds to an undetected $task2$. At the end of $task4$,  the capacitor voltages were measured to be $V_{C1} = \SI{2.79}{\volt}, V_{C2}  = \SI{2.80}{\volt}, V_{C3}  = \SI{2.80}{\volt}, V_{C4}  = \SI{2.79}{\volt}$.

In the second profile, we increased the duration of $task3$ so that it mimics an infected task. A similar procedure as in the case of the first profile was followed. In the beginning, the status of the capacitors were $V_{C1} = \SI{3.10}{\volt}, V_{C2}  = \SI{3.12}{\volt}, V_{C3}  = \SI{3.11}{\volt}, V_{C4}  = \SI{3.16}{\volt}$. As in the case of the regular profile, $task1$ was launched at instance $1$ and $task2$ was launched at instance $2$. However,  the execution of $task4$ was delayed by prolonging the execution of $task3$. This time, $task3$ was set to run for \SI{300}{\milli\second}. However, the MCU traced the load consumption in every $\SI{50}{\milli\second}$  and realized that the load is possibly running in an error state since the execution time exceeded the previously measured value. This is confirmed at instance $4$ and hence the supply to the load was ceased. The details presented in Fig.~\ref{fig_lmm_bug_icu} show that the disconnection of the load happened after \SI{200}{\milli\second} from the beginning of $task3$. After the load was disconnected, the status of the capacitors was assessed as $V_{C1} = \SI{3.09}{\volt}, V_{C2}  = \SI{3.13}{\volt}, V_{C3}  = \SI{3.04}{\volt}, V_{C4}  = \SI{3.17}{\volt}$. Therefore, 
it becomes apparent that the abnormal execution of $task3$ exclusively impacts  $C_{2}$. 

\section{Conclusion}
\label{section_conclusion}
We have presented a detailed discussion and evaluation of InfiniteEn, a multi-source energy harvesting platform for batteryless IoT devices. InfiniteEn is built on top of the re-configurable energy buffer concept. However, InfiniteEn differs from other existing architectures for its novel energy combiner cum energy rate sensor and a target agnostic LMM. The energy combiner achieves an efficiency of 88\%.  Apart from combining energy from multiple sources, the energy combiner circuit is found to be effective in estimating the harvester current and energy stored in the buffer. The LMM can detect any discharge current higher than  \SI{3}{\milli\ampere} from a \SI{15}{\milli\farad} supercapacitor and \SI{15}{\milli\ampere} or more from a \SI{100}{\milli\farad} capacitor. The minimum detectable current depends on the size of the supercapacitor, the gain of the differentiator and the reference voltage of the comparator. For a fixed gain and the reference voltage, the minimum detectable current increases with the capacity of the energy buffer. However, this is justifiable considering the fact that larger buffers have higher energy storage and hence a longer time to discharge.  Further, the response time of the LMM is observed to be less than \SI{11}{\milli\second} at slowly changing input voltages.   Using the inference provided by the LMM, InfiniteEn can manage its storage buffer and detect any abnormalities in the target's energy consumption.

\section*{Acknowledgment}
This research has received funding from  the European Union's Horizon 2020 research and innovation program under the Marie Sokolowski-Curie grant agreement No 813114.

\bibliographystyle{IEEEtran}
\bibliography{references.bib}

\end{document}